\documentclass{emulateapj}
\usepackage{color}

\slugcomment{To be submitted to ApJ\\ \indent DES-2017-0221 \\ \indent FERMILAB-PUB-17-300-AE}

\shorttitle{Hor I Chemical Analysis}
\shortauthors{Nagasawa et al.}

\begin{document}

\title{Chemical Abundance Analysis of Three $\alpha$-Poor, Metal-Poor Stars in the Ultra-Faint Dwarf Galaxy Horologium I\altaffilmark{*}}

\altaffiltext{*}{This paper includes data gathered with the 6.5m Magellan Telescopes located at Las Campanas Observatory, Chile. This paper also includes data based on observations made with the ESO Very Large Telescope at Paranal Observatory, Chile (ID 096.D-0967(B); PI: E. Balbinot)}


\author{D.~Q. Nagasawa\altaffilmark{1}, J.~L.~Marshall\altaffilmark{1}, J.~D.~Simon\altaffilmark{2}, T.~T. Hansen\altaffilmark{2}, T.~S.~Li\altaffilmark{3}, R.~A.~Bernstein\altaffilmark{2}, E.~Balbinot\altaffilmark{4}, A.~Drlica-Wagner\altaffilmark{3}, A.~B.~Pace\altaffilmark{1}, L.~E. Strigari\altaffilmark{1}, C.~M. Pellegrino\altaffilmark{1}, D.~L. DePoy\altaffilmark{1}, N.~B. Suntzeff\altaffilmark{1}, K.~Bechtol\altaffilmark{5}, T.~M.~C.~Abbott\altaffilmark{6}, F.~B.~Abdalla\altaffilmark{7,8}, S.~Allam\altaffilmark{3}, J.~Annis\altaffilmark{3}, A.~Benoit-L{\'e}vy\altaffilmark{9,7,10}, E.~Bertin\altaffilmark{9,10}, D.~Brooks\altaffilmark{7}, A.~Carnero~Rosell\altaffilmark{11,12}, M.~Carrasco~Kind\altaffilmark{13,14}, J.~Carretero\altaffilmark{15}, C.~E.~Cunha\altaffilmark{16}, C.~B.~D'Andrea\altaffilmark{17}, L.~N.~da Costa\altaffilmark{11,12}, C.~Davis\altaffilmark{16}, S.~Desai\altaffilmark{18}, P.~Doel\altaffilmark{7}, T.~F.~Eifler\altaffilmark{19,20}, B.~Flaugher\altaffilmark{3}, P.~Fosalba\altaffilmark{21}, J.~Frieman\altaffilmark{3,22}, J.~Garc\'ia-Bellido\altaffilmark{23}, E.~Gaztanaga\altaffilmark{21}, D.~W.~Gerdes\altaffilmark{24,25}, D.~Gruen\altaffilmark{16,26}, R.~A.~Gruendl\altaffilmark{13,14}, J.~Gschwend\altaffilmark{11,12}, G.~Gutierrez\altaffilmark{3}, W.~G.~Hartley\altaffilmark{7,27}, K.~Honscheid\altaffilmark{28.29}, D.~J.~James\altaffilmark{30}, T.~Jeltema\altaffilmark{31}, E.~Krause\altaffilmark{16}, K.~Kuehn\altaffilmark{32}, S.~Kuhlmann\altaffilmark{33}, N.~Kuropatkin\altaffilmark{3}, M.~March\altaffilmark{17}, R.~Miquel\altaffilmark{34,15}, B.~Nord\altaffilmark{3}, A.~Roodman\altaffilmark{16,26}, E.~Sanchez\altaffilmark{35}, B.~Santiago\altaffilmark{36,11}, V.~Scarpine\altaffilmark{3}, R.~Schindler\altaffilmark{26}, M.~Schubnell\altaffilmark{25}, I.~Sevilla-Noarbe\altaffilmark{35}, M.~Smith\altaffilmark{37}, R.~C.~Smith\altaffilmark{6}, M.~Soares-Santos\altaffilmark{3}, F.~Sobreira\altaffilmark{38,11}, E.~Suchyta\altaffilmark{39}, G.~Tarle\altaffilmark{25}, D.~Thomas\altaffilmark{40}, D.~L.~Tucker\altaffilmark{3}, A.~R.~Walker\altaffilmark{6}, R.~H.~Wechsler\altaffilmark{41,16,26}, R.~C.~Wolf\altaffilmark{17}, B.~Yanny\altaffilmark{3}}

\affil{\noindent $^1$Mitchell Institute for Fundamental Physics and Astronomy and Department of Physics and Astronomy, Texas A\&M University, College Station, TX 77843-4242, USA}
\affil{$^2$Observatories of the Carnegie Institution of Washington, 813 Santa Barbara St., Pasadena, CA 91101, USA}
\affil{$^3$Fermi  National  Accelerator  Laboratory, P. O. Box 500, Batavia, IL 60510, USA}
\affil{$^4$Department of Physics, University of Surrey, Guildford GU2 7XH, UK}
\affil{$^5$LSST, 933 North Cherry Avenue, Tucson, AZ 85721, USA}
\affil{$^{6}$Cerro Tololo Inter-American Observatory, National Optical Astronomy Observatory, Casilla 603, La Serena, Chile}
\affil{$^{7}$Department of Physics \& Astronomy, University College London, Gower Street, London, WC1E 6BT, UK}
\affil{$^{8}$Department of Physics and Electronics, Rhodes University, PO Box 94, Grahamstown, 6140, South Africa}
\affil{$^{9}$CNRS, UMR 7095, Institut d'Astrophysique de Paris, F-75014, Paris, France}
\affil{$^{10}$Sorbonne Universit\'es, UPMC Univ Paris 06, UMR 7095, Institut d'Astrophysique de Paris, F-75014, Paris, France}
\affil{$^{11}$Laborat\'orio Interinstitucional de e-Astronomia - LIneA, Rua Gal. Jos\'e Cristino 77, Rio de Janeiro, RJ - 20921-400, Brazil}
\affil{$^{12}$Observat\'orio Nacional, Rua Gal. Jos\'e Cristino 77, Rio de Janeiro, RJ - 20921-400, Brazil}
\affil{$^{13}$Department of Astronomy, University of Illinois, 1002 W. Green Street, Urbana, IL 61801, USA}
\affil{$^{14}$National Center for Supercomputing Applications, 1205 West Clark St., Urbana, IL 61801, USA}
\affil{$^{15}$Institut de F\'{\i}sica d'Altes Energies (IFAE), The Barcelona Institute of Science and Technology, Campus UAB, 08193 Bellaterra (Barcelona) Spain}
\affil{$^{16}$Kavli Institute for Particle Astrophysics \& Cosmology, P. O. Box 2450, Stanford University, Stanford, CA 94305, USA}
\affil{$^{17}$Department of Physics and Astronomy, University of Pennsylvania, Philadelphia, PA 19104, USA}
\affil{$^{18}$Department of Physics, IIT Hyderabad, Kandi, Telangana 502285, India}
\affil{$^{19}$Department of Physics, California Institute of Technology, Pasadena, CA 91125, USA}
\affil{$^{20}$Jet Propulsion Laboratory, California Institute of Technology, 4800 Oak Grove Dr., Pasadena, CA 91109, USA}
\affil{$^{21}$Institute of Space Sciences, IEEC-CSIC, Campus UAB, Carrer de Can Magrans, s/n,  08193 Barcelona, Spain}
\affil{$^{22}$Kavli Institute for Cosmological Physics, University of Chicago, Chicago, IL 60637, USA}
\affil{$^{23}$Instituto de Fisica Teorica UAM/CSIC, Universidad Autonoma de Madrid, 28049 Madrid, Spain}
\affil{$^{24}$Department of Astronomy, University of Michigan, Ann Arbor, MI 48109, USA}
\affil{$^{25}$Department of Physics, University of Michigan, Ann Arbor, MI 48109, USA}
\affil{$^{26}$SLAC National Accelerator Laboratory, Menlo Park, CA 94025, USA}
\affil{$^{27}$Department of Physics, ETH Zurich, Wolfgang-Pauli-Strasse 16, CH-8093 Zurich, Switzerland}
\affil{$^{28}$Center for Cosmology and Astro-Particle Physics, The Ohio State University, Columbus, OH 43210, USA}
\affil{$^{29}$Department of Physics, The Ohio State University, Columbus, OH 43210, USA}
\affil{$^{30}$Astronomy Department, University of Washington, Box 351580, Seattle, WA 98195, USA}
\affil{$^{31}$Santa Cruz Institute for Particle Physics, Santa Cruz, CA 95064, USA}
\affil{$^{32}$Australian Astronomical Observatory, North Ryde, NSW 2113, Australia}
\affil{$^{33}$Argonne National Laboratory, 9700 South Cass Avenue, Lemont, IL 60439, USA}
\affil{$^{34}$Instituci\'o Catalana de Recerca i Estudis Avan\c{c}ats, E-08010 Barcelona, Spain}
\affil{$^{35}$Centro de Investigaciones Energ\'eticas, Medioambientales y Tecnol\'ogicas (CIEMAT), Madrid, Spain}
\affil{$^{36}$Instituto de F\'\i sica, UFRGS, Caixa Postal 15051, Porto Alegre, RS - 91501-970, Brazil}
\affil{$^{37}$School of Physics and Astronomy, University of Southampton,  Southampton, SO17 1BJ, UK}
\affil{$^{38}$Instituto de F\'isica Gleb Wataghin, Universidade Estadual de Campinas, 13083-859, Campinas, SP, Brazil}
\affil{$^{39}$Computer Science and Mathematics Division, Oak Ridge National Laboratory, Oak Ridge, TN 37831}
\affil{$^{40}$Institute of Cosmology \& Gravitation, University of Portsmouth, Portsmouth, PO1 3FX, UK}
\affil{$^{41}$Department of Physics, Stanford University, 382 Via Pueblo Mall, Stanford, CA 94305, USA}

\email{dqnagasawa@physics.tamu.edu}

\begin{abstract}
We present chemical abundance measurements of three stars in the ultra-faint dwarf galaxy Horologium I, a Milky Way satellite discovered by the Dark Energy Survey. Using high resolution spectroscopic observations we measure the metallicity of the three stars as well as abundance ratios of several $\alpha$-elements, iron-peak elements, and neutron-capture elements. The abundance pattern is relatively consistent among all three stars, which have a low average metallicity of [Fe/H] $\sim -2.6$ and are not $\alpha$-enhanced ([$\alpha$/Fe] $\sim 0.0$). This result is unexpected when compared to other low-metallicity stars in the Galactic halo and other ultra-faint dwarfs and hints at an entirely different mechanism for the enrichment of Hor I compared to other satellites. We discuss possible scenarios that could lead to this observed nucleosynthetic signature including extended star formation, a Population III supernova, and a possible association with the Large Magellanic Cloud. 
\end{abstract}

\section{Introduction}

The past several decades have seen marked advancement in our understanding of how a galaxy like the Milky Way is assembled as well as how chemical enrichment processes could have evolved to produce the elements that now exist in the local Universe \citep[e.g.][]{bel13,frebelnorris15}.  From the early observational work of \cite{sz} a picture emerged that galaxies like the Milky Way most likely formed, at least in part, via hierarchical merging of smaller satellites.  Modern dark energy+cold dark matter ($\Lambda$CDM) N-body simulations of the Milky Way support this picture \citep[e.g.][]{bullock-johnston-MWassembly,robertson,johnston-MWchem}.  \\
\indent The past two decades have produced an abundance of new studies to compare to theory. Most of the progress in this field has been made through modern wide-field imaging surveys and subsequent spectroscopic study of the objects found in the survey images. For example, the Sloan Digital Sky Survey (SDSS) discovered many nearby ``ultra-faint'' dwarf galaxies in the Milky Way halo that have lower masses and higher mass-to-light ratios than previously known Milky Way satellites (see \citealt{mcconnachie} for a summary). More recently, new wide-field imaging surveys such as Pan-STARRS and the Dark Energy Survey \citep[DES;][]{des} have discovered even more Milky Way satellite galaxies. DES has been the most prolific of these surveys to date: the first two years of DES data alone have resulted in the discovery of 22 new candidate satellites located in and around the Milky Way halo \citep{bechtol_des,koposov_9,wagner_des,kimjerjen_hor2,kim,luque1,luque2}. Once discovered, these candidates must be confirmed through kinematics to be gravitationally-bound stellar associations via follow-up spectroscopic observations. Spectroscopic velocity measurements also yield a measure of the mass-to-light (M/L) ratio and a determination of whether a satellite is a dark matter-dominated dwarf galaxy or a baryon-dominated stellar cluster (see \citealt{galaxy} for a comprehensive definition).  The DES-discovered candidate satellites considered most likely to be ultra-faint dwarf galaxies have been selected for follow-up spectroscopy; five have subsequently been confirmed to be highly dark matter-dominated, low luminosity satellites: Reticulum II \citep[Ret II;][]{simon_ret2,koposov_ret2hor1}, Tucana II and Grus I \citep{walker_gru1}, Tucana III \citep{simon_tuc3}, Eridanus II \citep{li_eri2}, and Horologium I \citep[Hor I;][]{koposov_ret2hor1}, the last being the subject of this paper.  \\
\indent Due to their relative physical and therefore presumed chemical isolation at the time their stars were formed, ultra-faint dwarf galaxies provide opportunities to study not only the dark matter that dominates their mass profile but also the nucleosynthetic processes that occurred in the early Universe.  Star formation in these low-mass objects is likely to be highly influenced by only a few nucleosynthetic events \citep[e.g.][]{ji2015}.  And since star formation in ultra-faint dwarfs appears to have been quenched early in the history of the Universe, perhaps by reionization \citep{brown,wetzel,jeon}, a fossil record of the early star formation history of these objects is preserved today.\\
\indent Prior to the work presented here, three DES-discovered ultra-faint dwarfs have been the targets of detailed chemical study: Ret II, Tuc II, and Tuc III. In each of these galaxies a unique nucleosynthetic process is observed.  The majority of stars in Ret II that have been studied to date are so-called ``$r$-II'' stars, signifying that they show extreme enhancement in rapid neutron-capture elements \citep{ji-retII-rprocess,roederer-retII-chem}.  This nucleosynthetic signature can be explained by a single high-yield event (e.g. a binary neutron star merger or hypernova) polluting the gas cloud from which stars in the galaxy were still forming. The chemical diversity of stars in Tuc II is somewhat unlike that observed in previously studied ultra-faint dwarfs, and could be explained by a range of phenomena, not all of which follow the standard nucleosynthetic processes \citep{ji_tuc2}. \cite{hansen} report the discovery of a moderately $r$-process enhanced ($r$-I) star in Tuc III, a rare chemical signature when compared to the bulk of field stars in the Milky Way halo. The diverse abundance patterns observed in these galaxies, and the range of unusual phenomena invoked to explain them, suggests that star formation in the early Universe must have been a stochastic process that was highly variable on the mass scales of ultra-faint dwarf galaxies. If this trend holds for more of the newly discovered ultra-faint dwarfs, the study of chemical abundance patterns could provide an opportunity to improve our understanding of nucleosynthetic processes in the early Universe, in addition to providing further tests of the $\Lambda$CDM paradigm and the formation processes of galaxies like the Milky Way. \\ 
\indent In this paper we present a detailed chemical abundance analysis of the kinematically confirmed ultra-faint dwarf galaxy Hor I. Hor I is located at a heliocentric distance of 79 kpc, has a luminosity $M_V \sim -3.5\pm0.3$ mag \citep{bechtol_des}, and a mass-to-light ratio of $\sim 600$ \citep{koposov_ret2hor1}. The paper is organized as follows: in Section \ref{section:obs_and_data} we describe the observations and abundance analysis of three stars in Hor I; we present the abundance measurements in Section \ref{resultssection}. In Section \ref{section:discussion} we discuss the peculiar nature of the chemical abundance patterns observed in this galaxy. In Section \ref{section:concl} we conclude with a summary of the results and its impact.

\section{Observations and Data Analysis}
\label{section:obs_and_data}
\subsection{Observations and Data Reduction}

\begin{figure}[p]
\centering
\includegraphics[scale=.5]{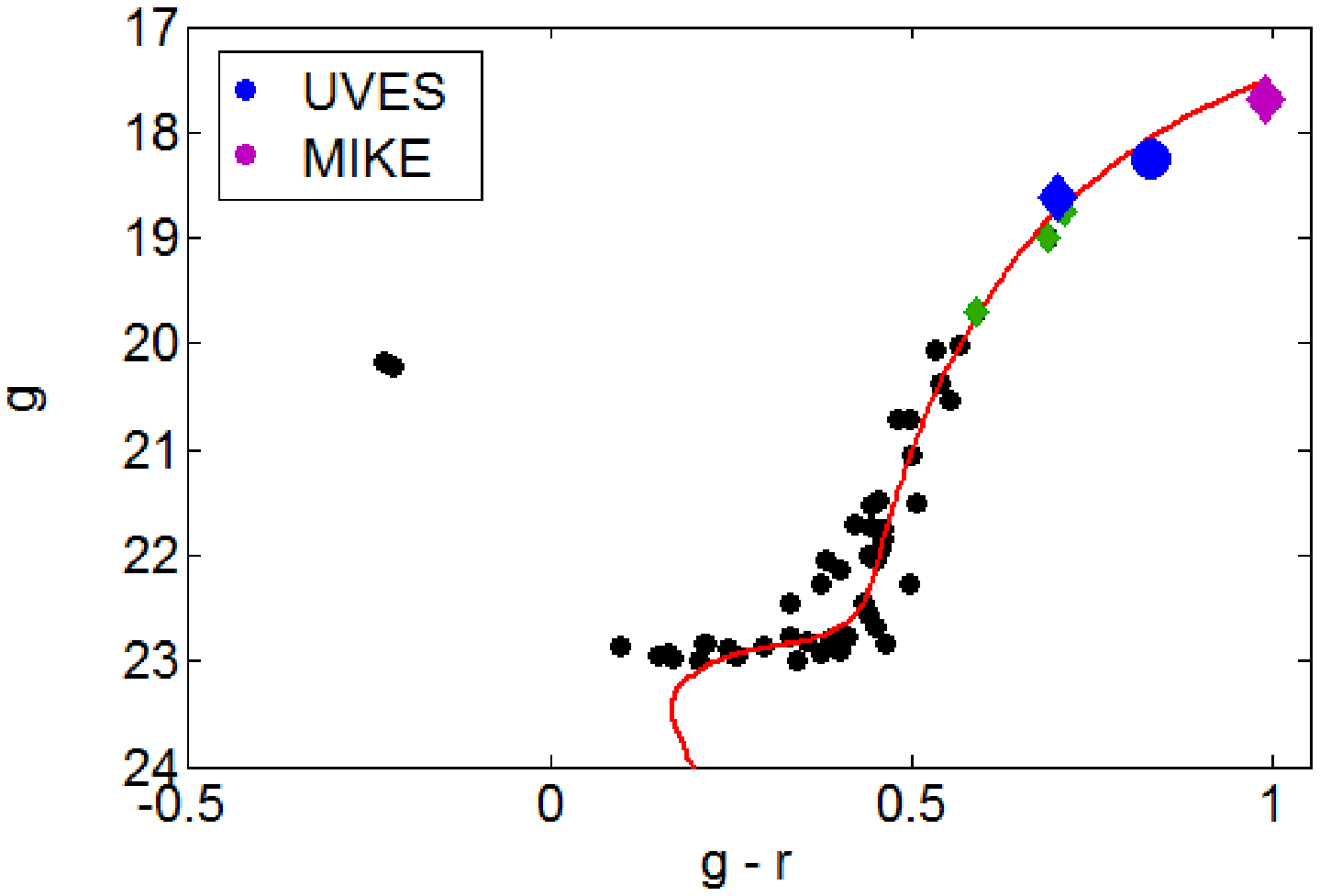}
\caption{Color-magnitude diagram of high probability ($> 70\%$) candidate member stars of Hor I from \citet{bechtol_des}. A Dartmouth isochrone \citep{dotter_isochrone} for a stellar population having $\tau = 12.5$ Gyrs, [Fe/H]=$-$2.5, [$\alpha$/Fe]=$+$0.0, and distance modulus $m - M = 19.7$ as derived by \citet{bechtol_des} is overplotted. The three stars studied in this work are indicated by larger points. The five diamond-shaped points are the confirmed member stars of \cite{koposov_ret2hor1}. Black points are unconfirmed member stars from \citet{bechtol_des}.}
\label{HR_diagram}
\end{figure}

\indent Observations were performed using the FLAMES-UVES spectrograph \citep{uves-instrument,flames} on the VLT in Paranal, Chile as part of program 096.D-0967(B) (PI: E. Balbinot) and the MIKE spectrograph \citep{bernstein-MIKE} at the Magellan-Clay Telescope at Las Campanas Observatory (PI: R. Bernstein).  In Figure \ref{HR_diagram} we present a color-magnitude diagram of the confirmed \citep{koposov_ret2hor1} and high-probability \citep{bechtol_des} member stars of Hor I, constructed using photometry from DES. DES astrometry and photometry of the three stars studied in this paper is presented in Table \ref{Observations}. \\
\subsubsection{UVES Observations}
\indent UVES observations took place on five nights over the months of December 2015 to January 2016 in fourteen 40-minute exposures. Stars were selected for UVES observation based on DES photometry, prior confirmation from \citet{koposov_ret2hor1}, and considerations related to fiber positioning due to simultaneous observations with the FLAMES-GIRAFFE spectrograph (Li et al. \emph{in prep.}).  Two stars were selected for UVES observations: DES J025540-540807, a confirmed member from previous observations using medium resolution spectra \citep{koposov_ret2hor1}, and DES J025543-544349, determined to be a likely member of Hor I \citep{bechtol_des}. Spectra of UVES targets were obtained in service mode. The 580 nm configuration was used, resulting in wavelength coverage of 4800 \AA\ $<\lambda<$ 6800 \AA\ with a $\sim$30 \AA\ gap in coverage around 5800 \AA\ due to the CCD chip gap and a spectral resolution of R$\sim$47,000. \\
\indent Bias subtraction, flat fielding, and spectral extraction were completed using the FLAMES-UVES Data Pipeline provided by the European Southern Observatories \citep{flames-uves-pipeline}. Due to the pixel oversampling (5 pixels per resolution element in the output spectrum) of the UVES spectra, we boxcar-smoothed the extracted spectra by 3 pixels in the wavelength dimension using the IRAF task \emph{boxcar}.\\
\indent Radial velocities were measured via Fourier cross-correlation of each exposure using the IRAF task \emph{fxcor} with a UVES spectrum of radial velocity standard HD140283 observed on a different night (29 May 2012) with the same instrument settings as our observations. We take the statistical error to be the standard deviation of the resulting velocities derived for each of the fourteen spectra, divided by the square root of the number of exposures (fourteen). A correction was applied based on the date of the observation to correct the radial velocities to the heliocentric frame. Each exposure was then shifted to rest wavelength and the fourteen spectra were mean-combined using 3-$\sigma$ rejection. \\
\indent We estimate the systematic error of the radial velocities as follows. All spectra for a single star obtained on a given night were median-combined and then Fourier cross-correlated with the combined spectra for the same star obtained on another night. To minimize the influence of noise, this cross-correlation was performed over the limited wavelength range of 5100 \AA\ $<\lambda<$ 5300 \AA\ centered on the strong Mg triplet lines. For DES J025540-540807, this night-to-night cross correlation yielded an average relative velocity of $0.51$ km s$^{-1}$ with respect to each other. For DES J025543-544349, the average relative velocity was $0.43$ km s$^{-1}$. \\
\indent The $S/N$ per resolution element of the two UVES spectra and measured radial velocities are presented in Table \ref{sn_and_dates}. The reported radial velocity error is the quadrature combination of the statistical and systematic errors. We note that the velocity of DES J025543-544349 is consistent with the other stars in Hor I, increasing the number of confirmed Hor I member stars from five to six.\\

\subsubsection{MIKE Observations}

\indent MIKE observations of DES J025535-540643, a confirmed Hor I member star \citep{koposov_ret2hor1}, took place on 06 August 2016 in five 30 minute  exposures. Using a 0.7 arcsec slit and 2$\times$2 pixel binning, the resulting spectrum has a resolution of R $\sim$ 22,000 ($\Delta\lambda = 0.23$ \AA) with coverage from 3310 \AA$<\lambda<$5000 \AA\ for the blue chip and 4825 \AA$<\lambda<$9150 \AA\ for the red chip. Reduction of the data, including bias correction, flat fielding, spectral extraction, wavelength calibration, and stacking were completed using the MIKE pipeline \citep{kelson-MIKE-pipeline}. \\
\indent For the spectrum obtained with MIKE, the radial velocity was measured by performing Fourier cross-correlation of the target star with a spectrum of radial velocity standard HD146051 \citep[radial velocity from][]{massarotti_RV} observed on the same night using the IRAF task \emph{fxcor}. A correction was applied based on the date of the observation to shift the radial velocities to the heliocentric frame.  Each spectral order was considered individually; the reported radial velocity is the average value of the velocity measured in each order and the reported error is the standard deviation of the radial velocities determined in each order of the spectrum. The measured $S/N$ per resolution element and radial velocity for DES J025535-540643 are presented in Table \ref{sn_and_dates}. \\

\subsection{Abundance Analysis}
\label{section:abundances}

We measured the equivalent widths of spectral features using the \emph{SPECTRE} program \citep{spectre-sneden}, with confirmation of the measurement of each line using the IRAF task \emph{splot}. 
The line list was generated from the Kurucz database \citep{kurucz-linelist} with updated laboratory transition probabilities from the NIST database \citep{nist}. Excitation potential, oscillator strength, and references for each line used in this analysis is listed in Table \ref{linelist}. For this analysis, it is assumed that these species are in local thermodynamic equilibrium (LTE). For CH and CN, we use dissociation energies of 3.47 eV \citep{masseronCH}  and 7.72 eV \citep{snedenCN} respectively.\\

\subsubsection{Determination of Stellar Parameters}
\label{eqw_analysis}
\indent Stellar parameters were derived spectroscopically from \ion{Fe}{1} and \ion{Fe}{2} lines using the \emph{abfind} package of the \emph{MOOG} program \citep{moog-sneden} and the $\alpha$-enhanced 1D plane-parallel Castelli-Kurucz model atmospheres \citep{castelli-kurucz-atmomodels}. We note that, although the stars studied here do not in fact turn out to be $\alpha$-enhanced, we choose to use the Kurucz $\alpha$-enhanced models for consistency with our previous and future work. From comparison tests using DES J025540-540807, which has an [Fe/H] $= -2.43$, we further note that at the lowest metallicities, the differences between the $\alpha$-enhanced and non-$\alpha$-enhanced Kurucz models are minimal, generally resulting in $\sim$0.05 dex additional error in the abundances (which is much smaller than our total adopted error). Using these models, we calculate an abundance for each \ion{Fe}{1} and \ion{Fe}{2} line individually. \\
\indent We take the mean abundance of all measured lines for each species to be the measured abundance and use the standard deviation of these abundances as a statistical error. The effective temperature was determined by iterating atmospheric models until there was no observed trend in calculated \ion{Fe}{1} abundance with excitation potential of the \ion{Fe}{1} lines. Surface gravity was determined by iterating until there was $1$-$\sigma$ agreement between abundances calculated for \ion{Fe}{1} and \ion{Fe}{2}. In several instances, \ion{Fe}{2} lines were measurable, but weak, which may contribute to a systematic error regarding the determined surface gravities. 
Microturbulence in the stellar atmosphere was determined by iterating microturbulent velocity until there was no observed trend in the calculated abundances of \ion{Fe}{1} with the  reduced equivalent width of the \ion{Fe}{1} lines. The same was done for \ion{Fe}{2} as well; the derived microturbulence for \ion{Fe}{2} was consistent with that derived for \ion{Fe}{1}.  Due to the known discrepancy between spectroscopically-derived and photometrically-derived effective temperature for metal-poor giant stars, a correction to the effective temperature was applied following \citet{frebel-metalpoor-measurement}. Surface gravity, microturbulence, and abundances were then recalculated. We determine the error in our stellar parameters by varying the stellar model and examining the resulting trends in excitation potential and reduced equivalent width. We calculate the final [Fe/H] of our stars from \ion{Fe}{1} due to the greater number of lines measured. Measured stellar parameters are presented in Table \ref{stellar_parameters}. \\
\subsubsection{Element Abundance Measurement Using Equivalent Widths}
\indent In both UVES and MIKE spectra, equivalent widths were measured for several species with strong, unblended absorption lines: \ion{Fe}{1}, \ion{Fe}{2}, \ion{Mg}{1}, and \ion{Ca}{1}. For \ion{Fe}{1} in particular, lines ranging across wavelength, excitation potential (E.P.), and transition probability log(\emph{gf}) were sampled in order to minimize systematic bias in abundance calculations. \\
\indent Due to the greater wavelength coverage of the MIKE spectrum, 60 \ion{Fe}{1} lines were measurable compared to the only 12 useful \ion{Fe}{1} in the UVES data. To ensure that the reduced number of lines in the UVES spectra would not systematically bias our measurements, the 12 \ion{Fe}{1} lines used in the UVES analysis were measured in the MIKE spectrum and analyzed separately from the full 60-line analysis. The difference between the two analyses in both stellar parameter determination and abundance measurement was within the uncertainties. We conclude then that the reduced number of lines in the UVES spectral analysis does not systematically affect the results. 
\subsubsection{Element Abundance Measurement using Synthetic Spectra}
\begin{figure*}[p]
\centering
\includegraphics[scale=.38]{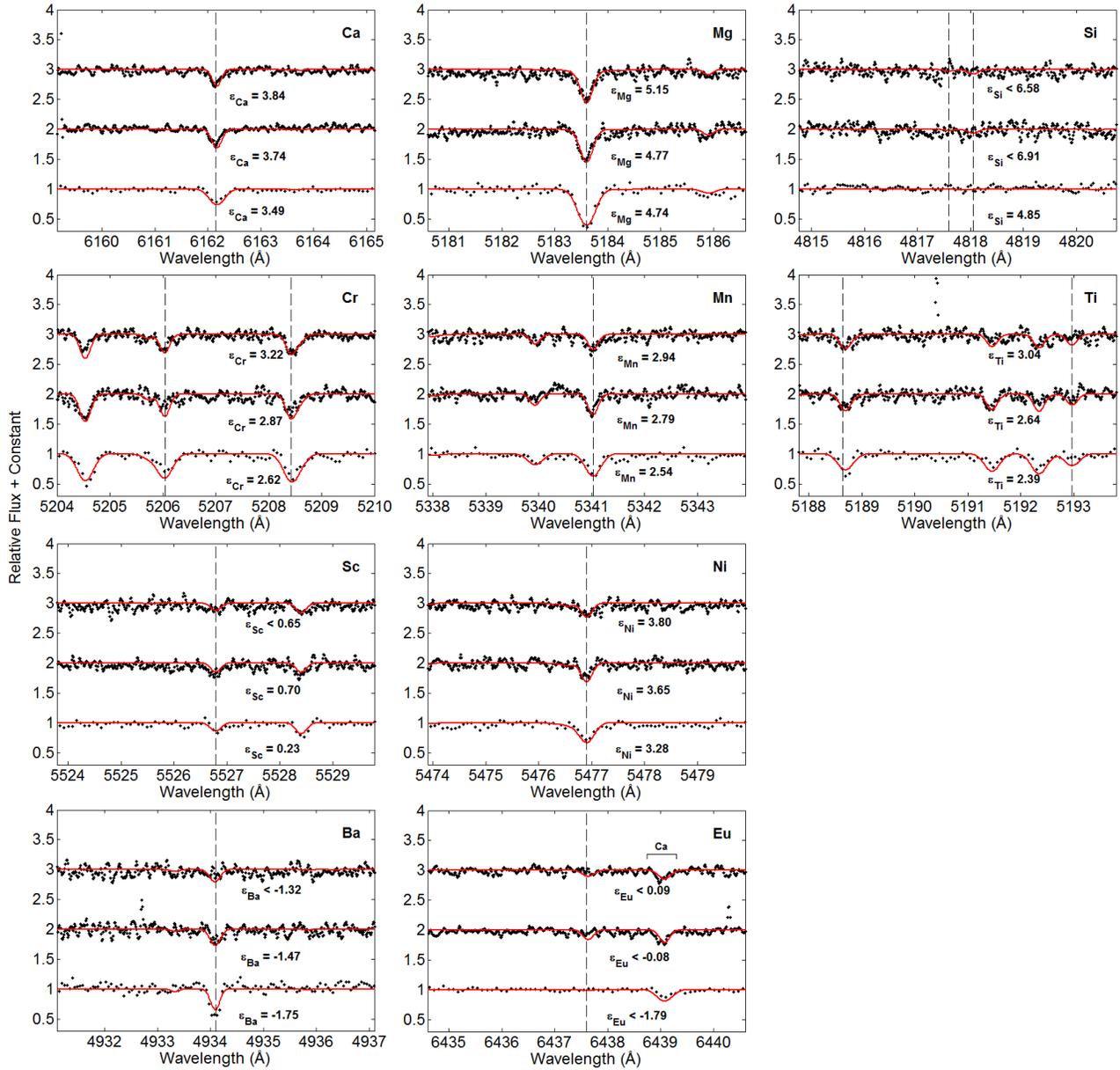}
\caption{Examples of synthetic spectra showing the region around the absorption features for Ca, Mg, Si, Cr, Mn, Ti, Sc, Ni, Ba, and Eu. In each panel, the top spectrum is DES J025540-540807, the middle spectrum is DES J025543-544349, and the bottom spectrum is DES J025535-540643. Observed data are plotted as black points, while synthetic spectra of the indicated $\epsilon_{X}$ are presented as red lines. Vertical dashed lines indicate the central wavelength of spectral features of the indicated element. }
\label{spectra}
\end{figure*}

\indent Spectral synthesis was done for elements that either did not have a large number of measurable lines due to low $S/N$ or due to blending and for elements where hyperfine structure and/or isotopic shifts needed to be considered. Using the stellar parameters derived, we have used spectral synthesis to measure the abundances of multiple elements in all three stars, specifically \ion{Si}{1}, \ion{Sc}{2}, \ion{Ti}{1}, \ion{Cr}{1}, \ion{Mn}{1}, \ion{Ni}{1}, \ion{Ba}{2}, and \ion{Eu}{2}. The increased wavelength coverage in the MIKE spectrum enables measurement of additional species in DES J025535-540643. For these measurements, multiple spectral lines were identified based on both their excitation potential and transition probability to be relatively strong (i.e., low excitation energies, high transition probabilities). Synthetic spectra were generated using the \emph{synth} package of the MOOG program \citep{moog-sneden} for a 40 \AA\ window centered on the line of interest. The abundances of Fe and Ca from equivalent width analysis were used as input in the synthesis. Spectra were generated varying the abundance of the elements of interest in [X/H] steps of 0.10--0.125 dex. A Gaussian function was utilized in the smoothing of the synthetic spectra, which was roughly what was expected based on spectrograph resolution. If available in the 40 \AA\ window, a \ion{Fe}{1} or \ion{Ca}{1} line was used to ensure that the Gaussian-smoothed synthetic spectrum using the equivalent width-derived stellar parameters was able to reproduce the observational data, generally reproducing observational data to $\sim0.10$ dex. Best fit spectra were selected by eye based on the $\chi^2$ minimization output in MOOG. Synthesis was also used to confirm the abundances derived using equivalent width analysis. Upper limits were derived by comparisons to synthetic spectra. Models of varying element abundances were generated until a model produced a clear detection that would have been distinguishable from noise but is undetected in the observed spectrum of the star. Sample synthetic spectra for elements measured using equivalent width analysis and spectral synthesis can be found in Figure \ref{spectra}, overlaid onto the observed spectra. \\

\indent Abundances are calculated as $\textrm{log}_{10}\left(\epsilon_{X}\right)$, which is defined in Equation \ref{epsilon_abundance} in terms of number density $N_X$. For reference, $\textrm{log}_{10}\left(\epsilon_{H}\right)$, where $N_H$ is the number density of hydrogen, is defined as 12. 
\begin{equation}
\textrm{log}_{10}\left(\epsilon_{X}\right) = \textrm{log}_{10}\left(\frac{N_X}{N_H}\right) + 12
\label{epsilon_abundance}
\end{equation}

\noindent Conversion into the more familiar [X/H] notation is performed using Equation \ref{dex_abundance} using measurements of $\textrm{log}_{10}\left(\epsilon_{X,\sun}\right)$ by \citet{asplund-solar}. Calculation of [X/Fe] is shown in Equation \ref{x/fe_abundance}.

\begin{equation}
\textrm{[X/H]}_\star = \textrm{log}_{10}\left(\epsilon_{X,\star}\right) - \textrm{log}_{10}\left(\epsilon_{X,\sun}\right)
\label{dex_abundance}
\end{equation}
\begin{equation}
\textrm{[X/Fe]}_\star = \textrm{[X/H]}_\star - \textrm{[Fe/H]}_\star
\label{x/fe_abundance}
\end{equation}

\noindent We present chemical abundance measurements in Table \ref{abundance_table}. We list each species measured, the number of lines measured for that species (N), ${log}_{10}(\epsilon_{X})$, metallicity, elemental abundance compared to iron, total error on the measurement (see discussion in Section \ref{error}), and method used to measure each species.  For lines for which we could only determine an upper limit, the total error was added to the value reported in the table, i.e. we attempt to report a conservative estimate of the upper limit.  For the UVES spectra we attempted to measure the abundances of several other elements, including Al, Co, Cu, Nd, Sr, Yb, and Zn, but could not obtain an upper limit lower than $+$4 dex for these elements due to the lack of strong lines in the UVES wavelength range.

\subsection{Error Analysis}
\label{error}
\indent In order to determine the uncertainty in the abundance measurements, we employ a method similar to \citet{mcwilliam_error} and account for the statistical and systematic errors separately. For lines measured using equivalent widths, we have calculated the mean abundance for multiple lines across excitation potential and transition probability space. We assume the standard deviation from this mean abundance represents our statistical error that arises from uncertainty in our equivalent width measurements. We take this to be the uncertainty on our abundance measurement for a single, unblended spectral feature. Therefore, by dividing by $\sqrt{N}$, where $N$ is the number of lines measured, we arrive at the statistical error in our abundance measurement that accounts for the multiple lines measured per element. \\
\indent To account for systematic errors introduced by the uncertainty in stellar parameter determination, we vary the stellar atmosphere model by the uncertainty in the stellar parameters individually. We then recalculate the abundance of each element using this perturbed model and determine the variation in our abundance measurement $\Delta$ log$_{10}$($\epsilon_{X}$) caused by the perturbation. We do this for effective temperature ($\pm 100$K), surface gravity ($\pm 0.2$ dex), and microturbulence ($\pm 0.5$ km s$^{-1}$).  The variation in abundance due to the perturbed stellar parameters is added in quadrature with the statistical error taken from the uncertainty in our equivalent width measurements, generating ${\Delta \textrm{log}_{10}\left(\epsilon_{X}\right)_{\textrm{,Total}}}$. \\
\indent For lines measured using spectral synthesis, we assess systematic errors as described above. However, because we use the consistency of multiple lines to measure element abundance, we cannot derive a statistical uncertainty in the same manner as the equivalent width analysis. We still remeasure abundances using a stellar atmosphere model perturbed by the uncertainty in the measured stellar parameters. Our perturbed model abundance is compared against the unperturbed abundance to determine the variation  $\Delta \epsilon$, which we take to be our systematic errors based upon the errors in our stellar parameter determination. We estimate, based on $S/N$ and the variations observed in our stellar parameter perturbation, that the statistical error associated with this measurement could be as high as $0.25$ dex. We therefore adopt this value as the statistical error for lines measured using synthetic spectra.  The final statistical error reported for lines measured using spectral synthesis is this value divided by the square root of the number of lines measured.  \\
\indent  We present the results of our error analysis in Table \ref{abundance_error_table}. The final errors reported in column 6 of Table \ref{abundance_table} and column 7 of Table \ref{abundance_error_table} are the quadrature sum of the systematic and statistical errors.\\

\section{Results}
\label{resultssection}
\begin{figure*}[p]
\centering
\includegraphics[scale=.36425]{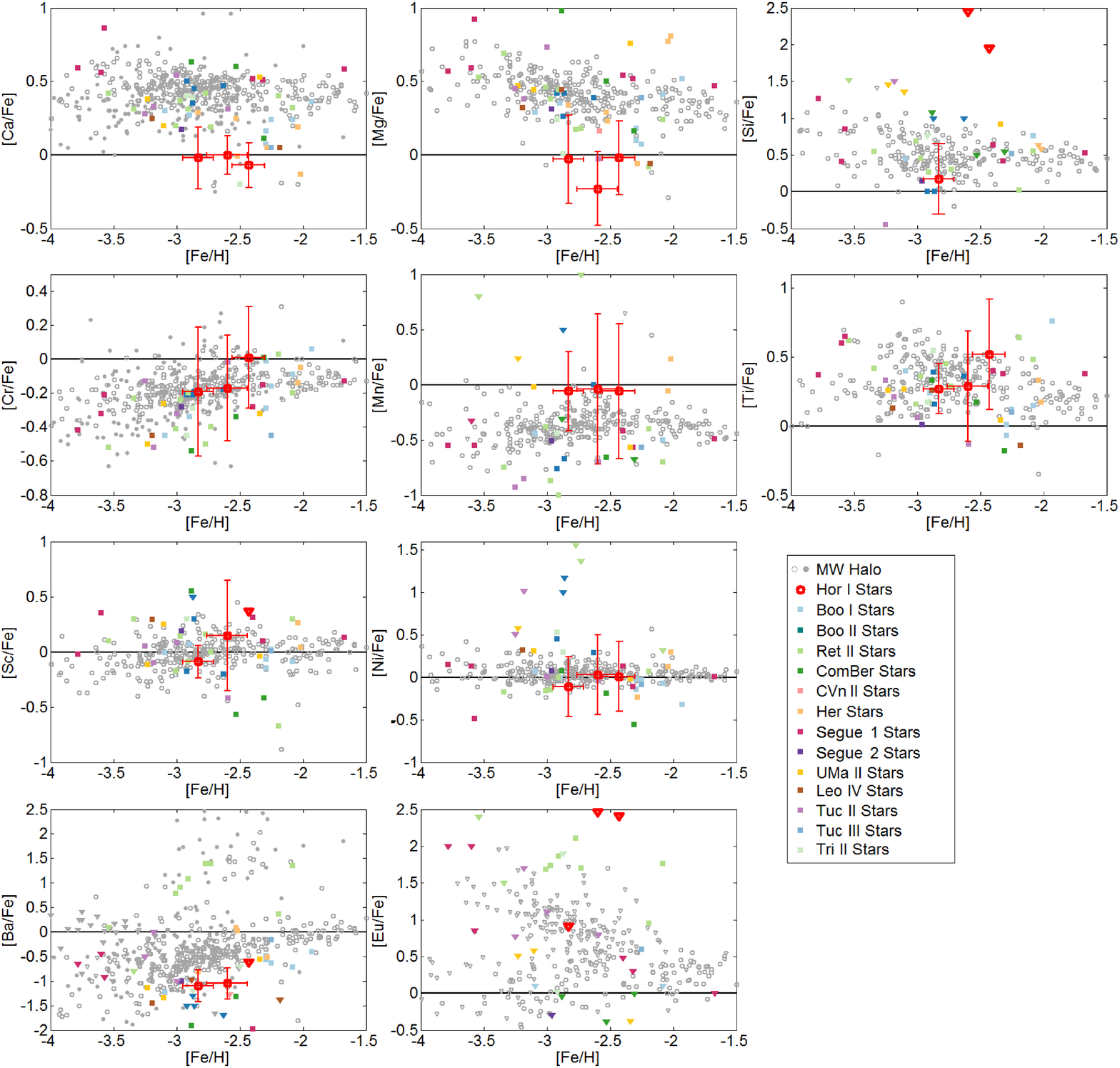}
\caption{Chemical abundance measurements of three Hor I member stars (red) compared to abundance measurements of stars in the ultra-faint dwarf galaxies Boo I \citep{norris_bootes1, ishigaki_bootes1, gilmore-bootesI-chem, frebel-bootes1-chem}, Boo II \citep{ji-bootesII-chem}, Ret II \citep{ji-retii-9star}, ComBer \citep{frebel_comber_uma2}, CVn II \citep{francois-her/cvn-chem}, Her \citep{koch_her0,koch_her, francois-her/cvn-chem}, Segue 1 \citep{frebel-segue1-chem}, Segue 2 \citep{roedererkirbysegue2}, UMa II \citep{frebel_comber_uma2}, Leo IV \citep{simon_leo4,francois-her/cvn-chem}, Tuc II \citep{ji_tuc2}, Tuc III \citep{hansen}, and Tri II \citep{venn_tri2,kirby_tri2} (various colored squares). Abundances of stars in the Milky Way halo from \citet{yong-halo} (filled gray) and \cite{roederer-halo} (open gray) are also shown. Error bars are shown only for the Hor I stars for clarity. Points denoted as $\bigtriangledown$ indicate an upper limit. The solar ratio ([X/Fe] = 0) is indicated by the solid black line.}
\label{star_comparison}
\end{figure*}

In Figure \ref{star_comparison} we compare Hor I stars to stars in the Milky Way halo and thirteen ultra-faint dwarf galaxies for which spectroscopic abundance analysis has been performed. The three Hor I stars are all of very low metallicity, ranging from $-2.83 <$ [Fe/H] $< -2.43$, and have similar $\alpha$-element and iron-peak element abundances. The measurement of Ba II in two stars and a consistent upper limit in the third star suggests that the abundance of neutron-capture elements in these three stars is also similar. \\ 
\indent In comparison to most other stars in the Milky Way halo and in other ultra-faint dwarf galaxies, the $\alpha$-element abundance of these three Hor I stars is low for their [Fe/H]. This can be seen in the [Ca/Fe] and [Mg/Fe] abundances. The detection of Si in one star, DES J025535-540643, is also consistent with the other $\alpha$-elements. There are a few stars in other ultra-faint dwarfs with similarly low [Ca/Fe] and [Mg/Fe] ($\sim 0$); however, these stars are generally more metal-rich, and no other ultra-faint dwarf has consistently low abundances for all $\alpha$-elements among all its measured member stars.  \\
\indent The iron-peak elements also present some unusual patterns. The abundances of Sc and Ni seem to be similar to that of stars in the halo and the other ultra-faint dwarfs. The abundance of Cr in Hor I is slightly elevated with respect to most other ultra-faint dwarfs, but still consistent with the abundances of halo stars. However, the abundance of Mn is $\sim 0.4$ dex higher than most halo stars and $\sim 0.6$ dex higher than the abundances found in other ultra-faint dwarfs. \\
\indent The abundance of Ba is similar to most other ultra-faint dwarfs. It does not present significant $s$-process or $r$-process enrichment like the stars in Ret II \citep{ji-retii-9star} or Tuc III \citep{hansen}. The upper limit of Eu found in DES J025535-540643 ([Eu/Fe] $< +0.91$) excludes it from being an $r$-II star (defined as [Eu/Fe] $> +1.0$), but does not exclude the possibility that it is an $r$-I star (defined as [Eu/Fe] $> +0.3$), where these definitions are taken from \cite{beers_christlieb_rhalo}. However, the low [Ba/Fe] of these three stars make it unlikely that they are $r$-process enhanced. \\
\indent Due to wavelength constraints, we could only measure C using the CH band in DES J025535-540643. Based upon the upper limit of [C/Fe] $< -0.14$, we can conclude that this star is not carbon-enhanced. 

\section{Discussion}
\label{section:discussion}
We discuss possible scenarios that could lead to the observed nucleosynthetic pattern of Hor I and compare Hor I stars to stars in the Milky Way with similar abundance patterns. In Section 4.1, we compare the abundance pattern of Hor I to stars found in the Milky Way halo with similar nucleosynthetic patterns. In Section 4.2, we discuss one plausible enrichment scenario, the early onset of Type Ia supernovae in Hor I. In Section 4.3, we compare the peculiar abundance pattern observed in Hor I to theoretical nucleosynthetic yield models. In Section 4.4, we discuss a possible association to the Large Magellanic Cloud as the cause of the abundance pattern measured in Hor I.
\subsection{Comparison with Similarly Peculiar Stars in the Milky Way Halo}
\indent The stars in Hor I are not the first metal-poor, $\alpha$-poor stars to be discovered.  For example, in a detailed chemical abundance study of stars found in a search for the most metal-poor stars in the Galactic halo, \citet{ivans} reported chemical abundance measurements of two additional low-$\alpha$, low-metallicity stars in the Galactic halo: G4-36 and CS 22966-043, and found that these two and BD +80$\degr$ 245 all have [Fe/H]$\sim$ $-$2 and [Ca/Fe] $\sim$ 0.5 dex below the mean halo value (\citealt{ivans} report [Ca/Fe] $=+0.31$ for the halo). Interestingly, these three stars also have iron-peak overabundances that are qualitatively similar to the Hor I stars, with BD +80$\degr$ 245 having the most similar abundances to our stars. \\
\indent Other studies have discovered extremely metal-poor stars having peculiar abundances: \cite{cohen} and \cite{haschke} report discoveries of extremely low-metallicity, low-$\alpha$ stars; \cite{caffau} found four extremely metal-poor stars ([Fe/H] $\sim$ $-$3.7) with even lower [$\alpha$/Fe] ratios than we measure in Hor I. Each of these studies invoke various theoretical supernova yield models to explain the observed abundance patterns, which are plausible explanations but in most cases do not perfectly match the observations.\\
\indent More recently, a metal-poor ([Fe/H]=$-$2.5) star having low-$\alpha$ enhancement ([$\alpha$/Fe]$\sim$ $-$0.4), SDSS J0018-0939, was discovered in the SDSS survey \citep{aoki}. The authors suggest that this star, whose observed abundance patterns are compared to theoretical nucleosynthetic yield models of a pair-instability supernova \citep[PISN;][]{heger-woosley-PISN}, may represent the first observational evidence of a PISN.\\
\indent \citet{simon_sculptor} found 2 stars in Sculptor with similar chemical abundances but at an average [Fe/H] of $\sim -3.9$, which is much more metal poor than Hor I. Scl 11\_1\_4296 had depleted abundances of Mg, Ca, and Si. Scl 07-50 had similarly low Ca and Si, but a Mg abundance that is consistent with the Milky Way halo. They concluded that these stars were the second generation of stars formed in the galaxy and that the chemical signatures were reproducible using Population III supernovae nucleosynthetic models. \\
\indent In an attempt to explain why previous observational searches for metal-free stars had largely failed, \cite{karlsson} constructed theoretical models for the early chemical enrichment of the Milky Way, showing that the lack of metal-free stars in the Galactic halo that are observable today is in fact expected if the first stars to form in the Universe were very massive \citep{bromm_massive_stars}.  In their model, the Galactic halo is assembled from stars formed during the assembly of ``atomic-cooling halos'' centered on minihalos each holding a Population III star. The model predicts that the earliest stars formed that are still observable today in fact should be very (not extremely) metal-poor stars, with [Fe/H]=$-$2.5 and low $\alpha$ abundances. They also show that stars with this chemical signature should be quite rare, about 1 star in 500 in the Galactic halo, though this may not be the case in ultra-faint dwarfs. 
This picture is consistent with hierarchical structure formation as well as, at least qualitatively, with the number of halo stars discovered to date having similarly peculiar abundance patterns.\\
\indent These previous discoveries of stars having similar observed abundance patterns to the Hor I stars studied here show that the abundance pattern we measure is not unique. However, it is quite interesting to find three very similar stars colocated in one low-mass galaxy. The halo stars described in the above studies are rare and unusual enough, when compared to other halo stars, to warrant special attention by those authors.  We suggest that those peculiar halo stars could have formed in small galaxies like Hor I, in which pollution by a single PISN occurred early in the star formation history of the galaxy (we investigate this in more detail in Section 4.3).  Those smaller satellites would then have been accreted into the Milky Way halo, leaving small numbers of halo stars with unusual abundance patterns sprinkled throughout the halo, as is observed. This scenario is consistent with the idea that the ultra-faint dwarfs are small contributors (by mass) to the accretion history of the Milky Way, as predicted by $\Lambda$CDM theory, and could perhaps be further confirmed if adequate numbers of similarly peculiar stars were found and their kinematic properties are consistent with having originated in the same accreted satellite. This last suggestion may be testable once Gaia proper motions are added to the measured radial velocities, enabling full position and kinematic information.\\

\subsection{Extended Star Formation in Hor I?}
\indent One plausible scenario that could explain the chemical abundances of Hor I is an early onset of Type Ia supernovae. In our current understanding of chemical evolution \citep{tinsley}, as a star-forming gas cloud collapses the most massive stars form early, quickly evolving to produce Type II supernovae and thereby enriching the surrounding gas cloud with the $\alpha$-elements O, Mg, Si, S, Ca and Ti. The next stars that form in this $\alpha$-rich environment would then be $\alpha$-enhanced stars with typical [$\alpha$/Fe] values $\geq 0.3$. As the stellar population continues to evolve, at some  later time Type Ia supernovae, which have characteristically low yields in $\alpha$-elements and greater yields of the iron-peak elements (Cr, Mn, Fe, Co, Cu), begin to dominate nucleosynthesis. The Type Ia supernovae then enrich the surrounding environment, thereby lowering the relative abundance of $\alpha$-elements and increasing the abundance of iron-peak elements. Stars formed after the transition between Type II-dominated nucleosynthesis and Type Ia-dominated nucleosynthesis would therefore present abundance ratios closer to the solar ratio ($\alpha$/Fe] $\sim 0$). This process produces a characteristic ``knee'' in the [$\alpha$/Fe] ratios across a range of metallicities, where metallicity, or [Fe/H], increases with time as the isolated stellar population enriches itself in iron.  In principle, the slope and the position of the knee can provide information about the rate and the time respectively at which this transition occurred in a given stellar population.  \cite{andy_araa} provides a comprehensive description of this story, which describes the observed abundances of stars in the Milky Way halo quite well.\\
\indent Presumably a similar series of events to that described above occurs in all stellar populations, where the specifics of the time delay, or, equivalently, metallicity, at which the transition between $\alpha$-rich to $\alpha$-poor star formation is determined by the star formation rate and initial mass function of the stellar population. This effect has been observed in dwarf galaxies using both detailed abundance measurements from high-resolution spectroscopy \citep[e.g.][]{venn2004,koch_car,hendricks} as well as with medium resolution spectroscopy \citep{kirby_halo}. The trend holds for lower mass objects as well: \cite{vargas-alpha} studied an ensemble of ultra-faint dwarf galaxies and determined that the transition between Type II- and Type Ia-dominated nucleosynthesis typically occurs in these objects at a ``time'' when [Fe/H] $\sim -2.3$, based on the summary properties of eight ultra-faint dwarfs. According to these results, stars in ultra-faint dwarfs that are more metal-poor than [Fe/H] $\sim -2.3$ generally should have formed in the $\alpha$-rich environment produced by Type II supernovae and thus present super-solar $\alpha$-element abundance. Conversely, stars with [Fe/H] $> -2.3$ were produced after Type Ia supernovae began to pollute the surrounding environment with iron-peak elements and would therefore show [$\alpha$/Fe] $\sim 0$. \\
\indent \cite{vargas-alpha} also determine that star formation in ultra-faint dwarfs occurs after a minimum time delay for the onset of Type Ia supernova of at least 100 Myr. This picture is consistent with other work that places limits on the star formation histories of ultra-faint dwarf galaxies: deep Hubble Space Telescope imaging and Keck spectroscopy of ultra-faint dwarf galaxies show that their stars were formed early, with roughly 80\% of stars having formed by 12.8 Gyr ago and 100\% of stars formed by 11.6 Gyr ago \citep{brown}. This duration is consistent with an early but extended star formation history that would conform to the standard process of chemical evolution in a stellar population. Furthermore, the picture that has emerged to describe star formation in ultra-faint dwarf galaxies is that star formation began quickly, in some cases in a single burst of star formation \citep{firstgalaxy}, and was soon quenched, possibly by reionization \citep[e.g.][]{brown,wetzel,jeon}, leaving the stars in the ultra-faint dwarfs as a fossil record of conditions in the early Universe.\\
\indent  If we presume a similar chemical evolution timeline for Hor I and use [Fe/H] as an age indicator, our measurements imply that the onset of Type Ia supernovae and the subsequent chemical enrichment of the surrounding gas would have had to occur relatively earlier in Hor I than in other ultra-faint dwarfs. The lack of $\alpha$-elements in even the most metal-poor star, DES J025535-540643 ([Fe/H] $= -2.8 \pm 0.2$ and  [Ca/Fe] $= -0.05 \pm 0.15$), implies that the transition from Type II supernovae-dominated nucleosynthesis and Type Ia supernovae-dominated nucleosynthesis had to occur at a time when the metallicity of Hor I was [Fe/H] $< -2.8$. This would represent a very early transition between nucleosynthesis dominated by Type II supernovae and nucleosynthesis dominated by Type Ia supernovae compared to other ultra-faint dwarfs.\\
\indent We do note that it is somewhat presumptuous to draw strong conclusions from a sample of three stars in a galaxy. Furthermore, at least one other ultra-faint dwarf galaxy has shown a spread in $\alpha$-enhancement at the low end of its metallicity range, Ursa Major I \citep[UMa I;][]{vargas-alpha}. The ten stars studied by \cite{vargas-alpha} span nearly two orders of magnitude in metallicity with a wide spread in $\alpha$-abundance at the lowest metallicity end, i.e. UMa I contains at least two metal-poor, $\alpha$-poor stars that could have chemical abundances similar to the Hor I stars.  Unfortunately, the moderate-resolution spectroscopy used by \cite{vargas-alpha} does not permit detailed abundance analysis of many elements.  It should be noted, however, that UMa I may not fit the canonical picture of stellar populations \citep{jeon}.\\
\indent Limits on the duration of star formation in Hor I could be placed if it were possible to study a larger sample of member stars chemically. According to the standard picture of chemical evolution described above, some of those stars would be older than the three studied here, should have [Fe/H] $< -2.8$, and should show $\alpha$-element enhancement consistent with the knee observed in other galaxies. Alternatively, a larger sample of stars could be studied with medium-resolution spectra using techniques such as those used by \cite{kirby}.\\

\subsection{Comparison to Supernova Yield Models}
 
\begin{figure*}[p]
\centering
\includegraphics[scale=.335]{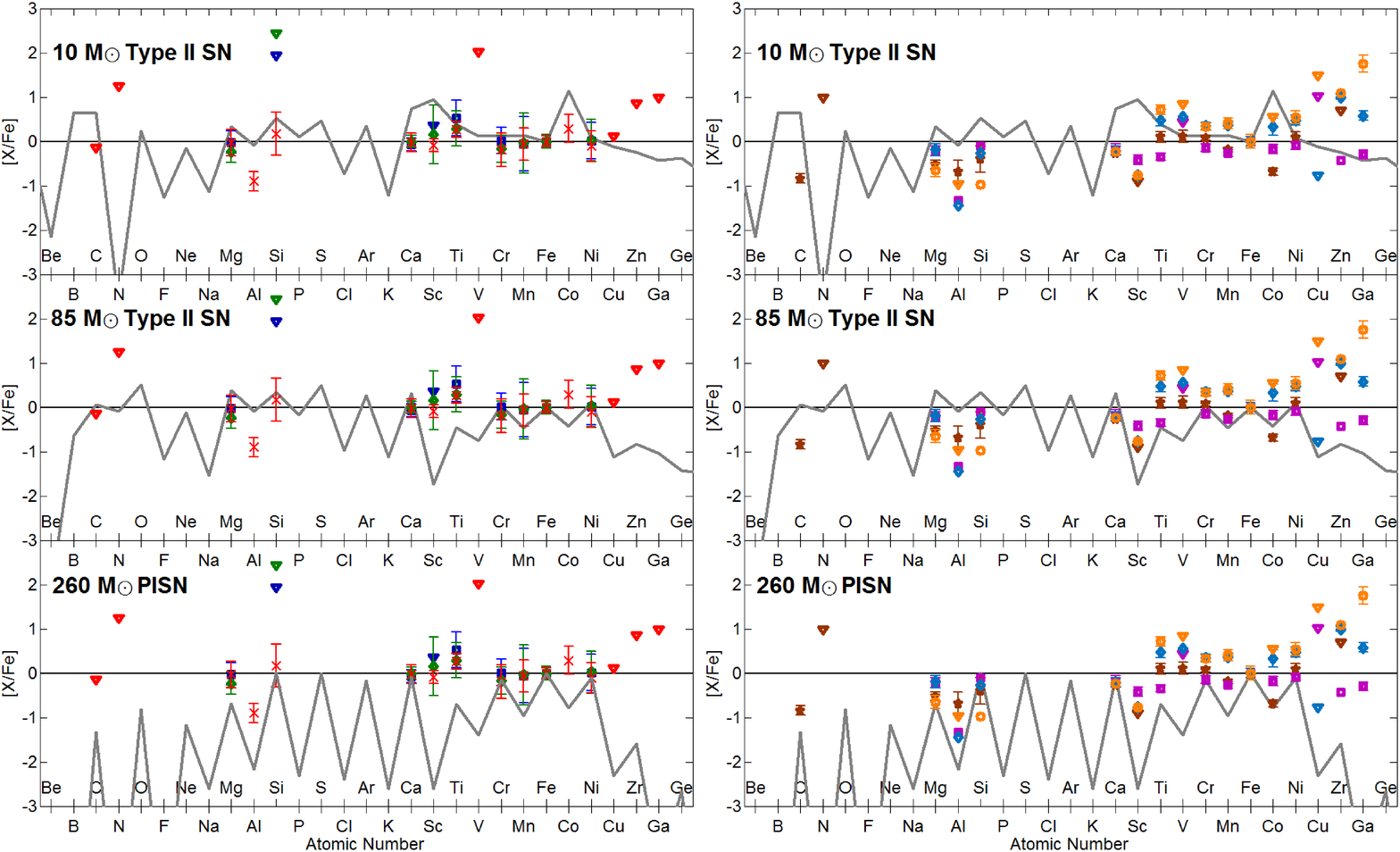}
\caption{Left: The three theoretical supernova yield models \citep{heger-woosley-PISN, heger-woosley-SN}  that best fit DES J025535-540643: a 10 $M_\sun$ Type II SN model (top), an 85 $M_\sun$ Type II supernova model (middle), and a 260 $M_\sun$ (130 $M_\sun$ He core) PISN model (bottom). For comparison, our measurements of [X/Fe] for all three stars are shown: DESJ025540-540807 (dark blue squares), DES J025543-544349 (green diamonds), DES J025535-540643 (red x's). Black lines indicate the solar ratio. Right: the same three supernova yield models with abundances of SDSS J0018-0939 \citep[brown stars;][]{aoki}, CS 22966-043 (pink squares), G4-36 (light blue diamonds), and BD +80$\degr$ 245 \citep[orange circles;][]{ivans}  shown for comparison. Points denoted as $\bigtriangledown$ indicate an upper limit. 
} 
\label{snyields}
\end{figure*}

\indent Another plausible scenario that could explain the peculiar abundance pattern we observe is that Hor I was host to a rare primordial supernova whose nucleosynthetic signature is preserved in the observable population of stars. \cite{firstgalaxy} suggest that the chemical signatures of low-mass ultra-faint dwarfs can be described by a single, long-lived, generation of stars that formed in the early Universe. In related work, \cite{ji2015} demonstrate that the chemical abundance patterns of these single events can be preserved in the second generation of stars. Though Hor I does not have the characteristically high $\alpha$-element abundance predicted by \citet{firstgalaxy} in their ``one-shot enrichment'' scenario, if \citet{aoki} are correct that their observed abundance patterns, which are similar to ours, are due to a PISN, then we expect that there must have been only a single nucleosynthetic event in Hor I. If there had been several generations of supernovae preceding the currently observed population, the peculiar abundance pattern produced in rare supernovae would be obscured by nucleosynthesis in other, more common Type II supernovae. By this reasoning, for the purposes of this analysis, we assume that the stars in Hor I are chemically primitive objects, and we explore the possibility that the observed abundances could be explained by the predicted yields of a single nucleosynthetic event. Therefore, in our comparison to nucleosynthetic yield models, we limit the number of events to a single Population III supernova that enriched the surrounding gas, creating the chemical abundance pattern observed today.\\ 
\indent To explore the possibility that the observed abundance pattern of Hor I may arise from a PISN, we have compared the abundances of DES J025535-540643 to various supernova yield models for Population III stars. These models can produce low [Ca/Fe] and [Mg/Fe] abundances, such as those observed in the three stars studied in Hor I. Since we were able to measure more elements in DES J025535-540643, we conduct this analysis only on this star. \\
\indent We used the STARFIT\footnote{See also http://starfit.org for routine and models} tool \citep[Chan et al. \emph{in prep.};][]{heger-woosley-SN} to compare our abundance measurements with Type II supernova nucleosynthetic yield models \citep[][and subsequent online updates in 2012]{heger-woosley-SN} for progenitors spanning a wide range in mass ($9.6$-$100 M_\sun$) and PISN nucleosynthetic yield models \citep{heger-woosley-PISN} for progenitors spanning a zero age main sequence (ZAMS) mass range of $140$-$260 M_\sun$. The STARFIT code calculates a $\chi^2$ statistic using abundance measurements and upper limits \citep[see][Equation 4]{heger-woosley-SN} and determines a best-fit supernova yield model. We used STARFIT to compare the observed abundance pattern of DES J025535-540643 against three categories of models; we present the parameters of the best fit models in Table \ref{sn_yield_model_table}. It should be noted that Sc and Cr are generally underproduced by yield models. \citet{heger-woosley-SN} assume that this is due to additional production sites that are unaccounted for and therefore discrepancies regarding these elements should be taken lightly. We therefore have STARFIT ignore them when fitting our abundance pattern. \\
\indent \citet{heger-woosley-SN} compute yields for non-rotating, metal-free Population III stars using initial Big Bang compositions from \citet{cybert-big-bang}. Due to the lack of a robust model for how a core-collapse supernova explodes, these computations utilize a piston model to simulate the explosion. \citet{heger-woosley-SN} compute nucleosynthetic yield models for two locations of the piston  (initial mass cut), one model for a piston at the base of the O shell ($S=4$ Piston Model) and one model for a piston at the edge of the Fe core ($Y_e$ Piston Model). We compare the abundance of DES J025535-540643 to both models using STARFIT. Using the model for a piston at the base of the O shell yields a best-fit model of a $10 M_\sun$ progenitor Type II supernova (mean squared residual $= 23.8$). Using the model for a piston at the edge of the Fe core yields a best-fit model of an $85 M_\sun$ progenitor Type II supernova (mean squared residual $= 28.2$).\\
\indent The explosion mechanism of a PISN is well-understood and is simulated to obtain theoretical nucleosynthetic yields by \citet{heger-woosley-PISN}. PISN progenitors enter a regime of electron/positron pair-production resulting in a collapse until O burning and Si burning produce enough energy to explode. This explosion results in low $\alpha$-element abundances, a low C abundance, and a strong odd-even effect. Comparison to the model nucleosynthetic yields of PISN using STARFIT gives a best-fit model of a $260 M_\sun$ ($130 M_\sun$ He core) PISN (mean squared residual $= 64.4$). It should be noted that this is the highest available PISN model used by STARFIT. It may be that the best fitting PISN model is beyond the available mass range. \\  
\indent In the left panels of Figure \ref{snyields}, we show the yield models that best fit DES J025535-540643 and the abundance measurements of all three stars observed in Hor I. As can be seen in the Figure, each model has difficulties in fitting the observed abundance patterns. The $10 M_\sun$ Type II supernova model produces too much C, Ca, Mg, and Co compared to our Hor I stars. These four elements produce contradictory requirements. The low upper limit on C and the abundances of Ca and Mg in DES J025535-540643 suggest that a higher energy explosion than provided by the $10 M_\sun$ Type II supernova is required, while the low Co abundance requires a lower energy explosion. The $85 M_\sun$ Type II supernova model produces too few iron-peak elements, implying that there is too much fallback and not enough iron-peak elements are synthesized and ejected. This model also does not produce enough Co, indicating that the energy of the explosion is too low. Finally, the $260 M_\sun$ ($130 M_\sun$ He core) PISN model produces a larger odd-even effect and a lower Co abundance than is observed in the stars of Hor I, which show essentially no odd-even effect. \\
\indent We compare BD +80$\degr$ 245, G4-36, CS 22966-043 \citep{ivans}, and SDSS J0018-0939 \citep{aoki} to the same models that best fit DES J025535-540643 in the right panels of Figure \ref{snyields}. It should be noted that the PISN model that we present is the same model suggested by \citet{aoki} as a possible fit for SDSS J0018-0939. For a common point of comparison for our best-fit models, we also used STARFIT to determine a best-fit PISN model for SDSS J0018-0939. The result was a best-fit model of a $260 M_\sun$ ($130 M_\sun$ He core) PISN (mean squared residual $= 159.6$). \citet{aoki} discussed the discrepancies in this PISN model fit to SDSS J0018-0939, specifically mentioning that the model predicts too much Si and too large of an odd-even effect for their observed abundance pattern. However, the model does fit their measured Co abundance. Our analysis of Hor I shares a similar problem in that the model's predicted odd-even effect is too large for our observed abundance pattern. The model does fit our Si abundance well, but underpredicts the amount of Co in DES J025535-540643, mirroring the discrepancies in SDSS J0018-0939. \\
\indent If it were possible to study a larger sample of member stars chemically then, if PISN were the underlying cause of the peculiarity in the observed abundance pattern of Hor I, the $\alpha$-element enhancement knee described previously would not be observed. It would require the chemical analysis of many more stars in Hor I to make any strong conclusions. \\

\subsection{Possible association with the LMC}
\indent An interesting question posed by the recent discovery of so many candidate ultra-faint dwarf galaxies in the outskirts of the Milky Way and located in the Southern hemisphere is whether they originated in the Milky Way or if rather they originated as satellites of satellites. Indeed, both groups announcing the discovery of Hor I \citep{bechtol_des, koposov_9}, as well as the kinematic confirmation work \citep{koposov_ret2hor1}, note Hor I's potential association with the Large Magellanic Cloud (LMC) due to its location and measured systemic velocity. 
Several recent theoretical studies have shown that the existence of satellites of satellites is predicted by simulations. Specifically, \cite{deason} use the ELVIS suite of N-body simulations to show that 2 to 4 of the 9 satellites discovered at the time that were found in close proximity to the LMC are expected to be associated with the LMC, while \cite{sales} use the Aquarius Project suite of zoomed-in cosmological simulations to show that 2 to 3 of all 46 dwarfs located within 300 kpc of the Milky Way should be associated with the LMC. Both of these works specifically state that Hor I has a high probability of being associated with the LMC according to their simulations. \cite{jethwa} use a complementary approach to these results and construct a dynamical model to determine which, if any, of the DES-discovered satellites could have Magellanic origins assuming the Milky Way--LMC system follows the distribution of sub-haloes predicted by $\Lambda$CDM.  Their model uses the satellites' observed positions and kinematic parameters to show that seven of the fourteen candidate DES satellites in the range $-$7 $<$ $M_V$ $<  - 1$ discovered by \cite{bechtol_des}, \cite{koposov_9}, and \cite{wagner_des} are likely to be satellites of the Large Magellanic Cloud (LMC) rather than of the Milky Way.  Their simulations produce predicted systemic velocities for the DES satellites, which must be confirmed by spectroscopic follow-up observations (only four of the fourteen had measured velocities at the time of publishing: Hor I, Ret II, Gru I, and Tuc II).  To date, of the satellites considered by \cite{jethwa}, Hor I's measured systemic velocity is by far the closest to the velocity predicted if Hor I were associated with the LMC.  \\
\indent If Hor I is indeed a satellite of the LMC, the chemical abundance pattern of Hor I could provide further interesting information about the relationship of the satellite to its host.  The LMC has an overall lower $\alpha$-enhancement than the Milky Way \citep[e.g.][]{pompeia, lapenna, lmcbar}. 
\cite{lmcbar} suggest that the lack of $\alpha$-elements implies a significantly different star formation history for the LMC than that of the Milky Way halo.  
Hence the lower $\alpha$-abundance of the Hor I stars may simply be due to its Magellanic origin, and the fact that early star formation in the LMC proceeded quite differently than in the halo of the Milky Way. The detailed abundance analysis of additional stars in Hor I, as well as of other candidate satellites of the LMC, would lend credence to this hypothesis. However, with only the three stars observed in this study, the chemical abundance pattern of Hor I does not exclude the possibility of an association with the LMC nor does it strongly suggest it. The strongest evidence that Hor I is a satellite of the LMC is the measured radial velocity of its member stars.

\section{Conclusions}
\label{section:concl}

We have measured the chemical abundances of three confirmed member stars in Hor I and have shown that it is yet another example of an ultra-faint dwarf galaxy having a peculiar abundance pattern. Hor I's average metallicity of [Fe/H] $\sim-2.6$ is not particularly exceptional, however, the observed $\alpha$ abundances are much lower than expected when compared to other metal-deficient stars. In addition, the abundances of other elements, in particular the iron-peak elements, present abundances close to the solar ratio, which is unusually high when compared to most Milky Way halo stars. We put forward the possibility that Hor I could have the earliest known transition between nucleosynthesis dominated by Type II supernovae and nucleosynthesis dominated by Type Ia supernovae. Alternatively, Hor I's chemistry could be explained by a PISN or it could be a satellite of the LMC.  In either case, our small sample of three stars is not enough to confirm these suggestions and additional member stars must be studied. \\
\indent Four DES-discovered ultra-faint dwarfs have been chemically studied in detail to date: Ret II \citep{ji-retII-rprocess,roederer-retII-chem}, Tuc II \citep{ji_tuc2}, Tuc III \citep{hansen}, and now Hor I.  In each case (with the possible exception of Tuc II), the brightest confirmed member stars show an unexpected and peculiar abundance pattern. Although a plausible explanation for the observed abundances can be invoked, the variety of explanations is large, suggesting that star formation processes in the early Universe may be highly stochastic.  These results suggest that study of additional ultra-faint dwarfs, and additional stars in these four previously studied ultra-faint dwarfs, may shed more light on how the first stars and galaxies were formed. However, probing the detailed chemical abundance patterns in many more confirmed member stars in Hor I will likely not be possible until the next generation of telescopes comes online in the next decade.

\acknowledgements{
\indent DQN wishes to express his most sincere and grateful thanks to Chris Sneden for extensive training in MOOG analysis, and to Katelyn Stringer for editorial comments. EB acknowledges financial support from the European Research Council (ERC-StG-335936).  C. Pellegrino was supported by NSF grant AST-1560223, ``REU Site: Astronomical Research and Instrumentation at Texas A\&M University.''\\
\indent Funding for the DES Projects has been provided by the U.S. Department of Energy, the U.S. National Science Foundation, the Ministry of Science and Education of Spain, the Science and Technology Facilities Council of the United Kingdom, the Higher Education Funding Council for England, the National Center for Supercomputing 
Applications at the University of Illinois at Urbana-Champaign, the Kavli Institute of Cosmological Physics at the University of Chicago, the Center for Cosmology and Astro-Particle Physics at the Ohio State University,
the Mitchell Institute for Fundamental Physics and Astronomy at Texas A\&M University, Financiadora de Estudos e Projetos, Funda{\c c}{\~a}o Carlos Chagas Filho de Amparo {\`a} Pesquisa do Estado do Rio de Janeiro, Conselho Nacional de Desenvolvimento Cient{\'i}fico e Tecnol{\'o}gico and the Minist{\'e}rio da Ci{\^e}ncia, Tecnologia e Inova{\c c}{\~a}o, the Deutsche Forschungsgemeinschaft and the Collaborating Institutions in the Dark Energy Survey.\\ 
\indent The Collaborating Institutions are Argonne National Laboratory, the University of California at Santa Cruz, the University of Cambridge, Centro de Investigaciones Energ{\'e}ticas, Medioambientales y Tecnol{\'o}gicas-Madrid, the University of Chicago, University College London, the DES-Brazil Consortium, the University of Edinburgh, 
the Eidgen{\"o}ssische Technische Hochschule (ETH) Z{\"u}rich, Fermi National Accelerator Laboratory, the University of Illinois at Urbana-Champaign, the Institut de Ci{\`e}ncies de l'Espai (IEEC/CSIC), the Institut de F{\'i}sica d'Altes Energies, Lawrence Berkeley National Laboratory, the Ludwig-Maximilians Universit{\"a}t M{\"u}nchen and the associated Excellence Cluster Universe, the University of Michigan, the National Optical Astronomy Observatory, the University of Nottingham, The Ohio State University, the University of Pennsylvania, the University of Portsmouth, 
SLAC National Accelerator Laboratory, Stanford University, the University of Sussex, Texas A\&M University, and the OzDES Membership Consortium. \\
\indent Based in part on observations at Cerro Tololo Inter-American Observatory, National Optical Astronomy Observatory, which is operated by the Association of Universities for Research in Astronomy (AURA) under a cooperative agreement with the National Science Foundation. \\
\indent The DES data management system is supported by the National Science Foundation under Grant Numbers AST-1138766 and AST-1536171. The DES participants from Spanish institutions are partially supported by MINECO under grants AYA2015-71825, ESP2015-88861, FPA2015-68048, SEV-2012-0234, SEV-2016-0597, and MDM-2015-0509, 
some of which include ERDF funds from the European Union. IFAE is partially funded by the CERCA program of the Generalitat de Catalunya. Research leading to these results has received funding from the European Research
Council under the European Union's Seventh Framework Program (FP7/2007-2013) including ERC grant agreements 240672, 291329, and 306478. We  acknowledge support from the Australian Research Council Centre of Excellence for All-sky Astrophysics (CAASTRO), through project number CE110001020. \\
\indent This manuscript has been authored by Fermi Research Alliance, LLC under Contract No. DE-AC02-07CH11359 with the U.S. Department of Energy, Office of Science, Office of High Energy Physics. The United States Government retains and the publisher, by accepting the article for publication, acknowledges that the United States Government retains a non-exclusive, paid-up, irrevocable, world-wide license to publish or reproduce the published form of this manuscript, or allow others to do so, for United States Government purposes. \\
\indent This paper has gone through internal review by the DES collaboration.\\
}

\begin{deluxetable*}{lccccccc}
\tablecaption{DES Astrometry and Photometry of Three Member Stars of Hor I} 
\tablehead{\colhead{ID} & \colhead{R.A. (2000)} & \colhead{Dec. (2000)} &  \colhead{$g$} & \colhead{$r$} & \colhead{$i$}& \colhead{$z$} & \colhead{$Y$}\\
&(deg)&(deg)&&
}
DES J025540-540807\tablenotemark{a} & $43.91793$ & $-54.13534$ & $18.67$ & $17.94$  & $17.67$ & $17.51$ & $17.50$ \\
DES J025543-544349 & $43.93246$ & $-54.08878$ & $18.30$ & $17.45$ & $17.14$ & $16.96$ & $16.93$ \\
DES J025535-540643\tablenotemark{b} & $43.89665$ & $-54.11222$ & $17.73$\tablenotemark{c} & $16.71$ & $16.35$ & $16.14$ & $16.10$ \\
\tablenotetext{a}{Referred to as Horo 9 by \citet{koposov_ret2hor1}.}
\tablenotetext{b}{Referred to as Horo 10 by \citet{koposov_ret2hor1}.}
\tablenotetext{c}{Note that \citet{koposov_ret2hor1} report g=19.31 mag for this star.}
\label{Observations}
\end{deluxetable*} 

\begin{deluxetable}{lcccc}
\tablecaption{Observing details} 
\tablehead{\colhead{ID} & \colhead{Instr.} & \colhead{$S/N$ } & \colhead{$S/N$} & \colhead{V$_\textrm{helio}$} \\ 
&&at 5300 \AA & at 6300 \AA & (km s$^{-1}$) }
DES J025540-540807 & UVES &  30 & 40 & $118.6 \pm 0.6$ \\
DES J025543-544349 & UVES & 35 & 40 & $114.3 \pm 0.5$\\
DES J025535-540643 & MIKE &  20 & 20 & $116.9 \pm 0.5$\\
\label{sn_and_dates}
\end{deluxetable}

\begin{deluxetable}{lcccl}
\tablecaption{Atomic Line Data}
\tablehead{\colhead{Species} & \colhead{$\lambda$} & \colhead{E.P.} & \colhead{log(\emph{gf}) } & \colhead{Reference}  \\ 
&(\AA)&(eV)&(dex)&}
\ion{Fe}{1}	&$	4045.81	$&$	1.48	$&\phs$	0.28	$&	\citet{kurucz-linelist}	\\
\ion{Fe}{1}	&$	4063.59	$&$	1.56	$&\phs$	0.06	$&	\citet{nist}	\\
\ion{Fe}{1}	&$	4071.74	$&$	1.61	$&$	-0.02	$&	\citet{kurucz-linelist}	\\
\ion{Fe}{1}	&$	4147.67	$&$	1.48	$&$	-2.10	$&	\citet{kurucz-linelist}	\\
\ion{Fe}{1}	&$	4216.18	$&$	0.00	$&$	-3.36	$&	\citet{kurucz-linelist}	\\
\ion{Fe}{1}	&$	4250.13	$&$	2.47	$&$	-0.41	$&	\citet{nist}	\\
\ion{Fe}{1}	&$	4260.47	$&$	2.40	$&\phs$	0.08	$&	\citet{nist}	\\
 \ion{Fe}{1}	&$	4415.12	$&$	1.61	$&$	-0.62	$&	\citet{kurucz-linelist}	\\
 \ion{Fe}{1}	&$	4427.31	$&$	0.05	$&$	-3.04	$&	\citet{kurucz-linelist}	\\
 $\vdots	$&$	\vdots	$&$	\vdots	$&$	\vdots	$&$	\vdots$ \\	
\ion{Tb}{2}	&$	4002.57	$&$	0.64	$&$	-0.49	$&	\citet{lawler_Tb,lawler_Tb_hfs}	\\
\ion{Tb}{2}	&$	4005.47	$&$	0.13	$&$	-0.02	$&	\citet{lawler_Tb}	\\
\ion{Tb}{2}	&$	4752.53	$&$	0.00	$&$	-0.55	$&	\citet{lawler_Tb}	\\
\ion{Dy}{2}	&$	3944.68	$&$	0.00	$&\phs$	0.11	$&	\citet{wickliffe_dy}	\\
\ion{Dy}{2}	&$	4103.31	$&$	0.10	$&$	-0.38	$&	\citet{wickliffe_dy}	\\
\ion{Dy}{2}	&$	4449.70	$&$	0.00	$&$	-1.03	$&	\citet{wickliffe_dy}	\\
\ion{Er}{2}	&$	3896.23	$&$	0.06	$&$	-0.12	$&	\citet{lawler_er}	\\
\ion{Er}{2}	&$	3938.63	$&$	0.00	$&$	-0.52	$&	\citet{kurucz-linelist}	\\
\ion{Th}{2}	&$	4019.13	$&$	0.00	$&$	-0.65	$&	\citet{kurucz-linelist}	\\

\label{linelist}
\tablecomments{This table is available in its entirety in machine-readable form.}
\end{deluxetable}

\begin{deluxetable*}{lccccc}
\tablecaption{Measured Stellar Parameters}
\tablehead{\colhead{ID} & \colhead{$T_\textrm{eff}$} & \colhead{log(g)} & \colhead{v$_\textrm{micro}$} & \colhead{[Fe/H]} & \colhead{[Ca/Fe]} \\
& (K) & (dex) & (km s$^{-1}$) & (dex)& (dex)
}
DES J025540-540807 & $5000 \pm 100$ & $2.0 \pm 0.2$ & $0.8 \pm 0.5$ & $-2.43 \pm 0.13$ & $-0.07 \pm 0.15$ \\
DES J025543-544349 & $4800 \pm 100$ & $1.5 \pm 0.2$ & $1.8 \pm 0.5$ & $-2.60 \pm 0.16$ & $+0.00 \pm  0.13$\\
DES J025535-540643 & $4500 \pm 100$ & $1.4 \pm 0.2$ & $3.5 \pm 0.5$ & $-2.83 \pm 0.12$ & $-0.02 \pm 0.21$ \\
\label{stellar_parameters}
\end{deluxetable*}

\begin{deluxetable}{lcccccl}
\tablecaption{Abundances of Three Confirmed Member Stars of Hor I}
\fontsize{8}{6}\selectfont
\tablehead{\colhead{Species} & \colhead{N} & \colhead{$\textrm{log}_{10}\left(\epsilon_{X}\right)$} & \colhead{[X/H]} & \colhead{[X/Fe]} & \colhead{Error} & \colhead{Method}}
\multicolumn{7}{c}{\rule{0pt}{1ex} DES J025540-540807 \rule{0pt}{1ex}}\\
\tableline
\ion{Mg}{1}	&$	2 	$&$		5.15	$&$		-2.45	$&$		-0.02	$&$	0.25	$&	Eq. Width	\\
\ion{Si}{1}	&$	4	$&$	<	6.58	$&$	<	-0.93	$&$	<	+1.95	$&$	0.45	$&	Spec. Synthesis	\\
\ion{Ca}{1}	&$	4	$&$		3.84	$&$		-2.50	$&$		-0.07	$&$	0.15	$&	Eq. Width	\\
\ion{Sc}{2}	&$	1	$&$	<	0.65	$&$	<	-2.50	$&$	<	+0.37	$&$	0.44	$&	Spec. Synthesis	\\
\ion{Ti}{1}	&$	3	$&$		3.04	$&$		-1.91	$&$		+0.52	$&$	0.40	$&	Spec. Synthesis	\\
\ion{Cr}{1}	&$	8	$&$		3.22	$&$		-2.42	$&$		+0.01	$&$	0.30	$&	Spec. Synthesis	\\
\ion{Mn}{1}	&$	3	$&$		2.94	$&$		-2.49	$&$		-0.06	$&$	0.61	$&	Spec. Synthesis	\\
\ion{Fe}{1}	&$	12	$&$		5.07	$&$		-2.43	$&$		+0.00	$&$	0.13	$&	Eq. Width	\\
\ion{Fe}{2}	&$	4	$&$		4.93	$&$		-2.57	$&$		-0.14	$&$	0.11	$&	Eq. Width	\\
\ion{Ni}{1}	&$	2	$&$		3.80	$&$		-2.42	$&$		+0.01	$&$	0.41	$&	Spec. Synthesis	\\
\ion{Ba}{2}	&$	3	$&$	<	-1.32	$&$	<	-3.50	$&$	<	-0.61	$&$	0.46	$&	Spec. Synthesis	\\
\ion{Eu}{2}	&$	2	$&$	<	0.09	$&$	<	-0.43	$&$	<	+2.41	$&$	0.41	$&	Spec. Synthesis	\\
\tableline
\multicolumn{7}{c}{\rule{0pt}{2ex}DES J025543-544349}\\
\tableline
\ion{Mg}{1}	&$	3	$&$		4.77	$&$		-2.83	$&$		-0.23	$&$	0.25	$&	Eq. Width	\\
\ion{Si}{1}	&$	4	$&$	<	6.91	$&$	<	-0.60	$&$	<	+2.45	$&$	0.45	$&	Spec. Synthesis	\\
\ion{Ca}{1}	&$	3	$&$		3.74	$&$		-2.60	$&$		+0.00	$&$	0.13	$&	Eq. Width	\\
\ion{Sc}{2}	&$	1	$&$		0.70	$&$		-2.45	$&$		+0.15	$&$	0.50	$&	Spec. Synthesis	\\
\ion{Ti}{1}	&$	3	$&$		2.64	$&$		-2.31	$&$		+0.29	$&$	0.40	$&	Spec. Synthesis	\\
\ion{Cr}{1}	&$	8	$&$		2.87	$&$		-2.77	$&$		-0.17	$&$	0.31	$&	Spec. Synthesis	\\
\ion{Mn}{1}	&$	3	$&$		2.79	$&$		-2.64	$&$		-0.04	$&$	0.68	$&	Spec. Synthesis	\\
\ion{Fe}{1}	&$	12	$&$		4.90	$&$		-2.60	$&$		+0.00	$&$	0.16	$&	Eq. Width	\\
\ion{Fe}{2}	&$	4	$&$		4.78	$&$		-2.72	$&$		-0.12	$&$	0.11	$&	Eq. Width	\\
\ion{Ni}{1}	&$	2	$&$		3.65	$&$		-2.57	$&$		+0.03	$&$	0.47	$&	Spec. Synthesis	\\
\ion{Ba}{2}	&$	3	$&$		-1.47	$&$		-3.65	$&$		-1.05	$&$	0.32	$&	Spec. Synthesis	\\
\ion{Eu}{2}	&$	2	$&$	<	-0.08	$&$	<	-0.60	$&$	<	+2.47	$&$	0.47	$&	Spec. Synthesis	\\
\tableline
\multicolumn{7}{c}{\rule{0pt}{2ex}DES J025535-540643}\\
\tableline
\ion{C}{0} (CH)	&$	1	$&$	<	5.10	$&$	<	-3.33	$&$	<	-0.14	$&$	0.36	$&	Spec. Synthesis	\\
\ion{N}{0} (CN)	&$	1	$&$	<	5.75	$&$	<	-2.08	$&$	<	+1.25	$&$	0.50	$&	Spec. Synthesis	\\
\ion{Mg}{1}	&$	4	$&$		4.74	$&$		-2.86	$&$		-0.03	$&$	0.30	$&	Eq. Width	\\
\ion{Al}{1}	&$	2	$&$		2.72	$&$		-3.73	$&$		-0.90	$&$	0.22	$&	Spec. Synthesis	\\
\ion{Si}{1}	&$	1	$&$		4.85	$&$		-2.66	$&$		+0.17	$&$	0.48	$&	Spec. Synthesis	\\
\ion{Ca}{1}	&$	4	$&$		3.49	$&$		-2.85	$&$		-0.02	$&$	0.21	$&	Eq. Width	\\
\ion{Sc}{2}	&$	3	$&$		0.23	$&$		-2.92	$&$		-0.09	$&$	0.15	$&	Spec. Synthesis	\\
\ion{Ti}{1}	&$	3	$&$		2.39	$&$		-2.56	$&$		+0.27	$&$	0.18	$&	Spec. Synthesis	\\
\ion{V}{1}	&$	1	$&$	 	1.80	$&$		-2.13	$&$		+0.70	$&$	0.30	$&	Spec. Synthesis	\\
\ion{Cr}{1}	&$	5	$&$		2.62	$&$		-3.02	$&$		-0.19	$&$	0.38	$&	Spec. Synthesis	\\
\ion{Mn}{1}	&$	1	$&$		2.54	$&$		-2.89	$&$		-0.06	$&$	0.36	$&	Spec. Synthesis	\\
\ion{Fe}{1}	&$	60	$&$		4.67	$&$		-2.83	$&$		+0.00	$&$	0.12	$&	Eq. Width	\\
\ion{Fe}{2}	&$	4	$&$		4.56	$&$		-2.94	$&$		-0.11	$&$	0.19	$&	Eq. Width	\\
\ion{Co}{1}	&$	3	$&$		2.45	$&$		-2.54	$&$		+0.29	$&$	0.32	$&	Spec. Synthesis	\\
\ion{Ni}{1}	&$	3	$&$		3.28	$&$		-2.94	$&$		-0.11	$&$	0.35	$&	Spec. Synthesis	\\
\ion{Cu}{1}	&$	3	$&$	<	1.16	$&$	<	-3.03	$&$	<	+0.12	$&$	0.32	$&	Spec. Synthesis	\\
\ion{Zn}{1}	&$	2	$&$	<	2.30	$&$	<	-2.26	$&$	<	+0.87	$&$	0.30	$&	Spec. Synthesis	\\
\ion{Ga}{1}	&$	1	$&$	<	0.78	$&$	<	-2.26	$&$	<	+1.00	$&$	0.43	$&	Spec. Synthesis	\\
\ion{Rb}{1}	&$	2	$&$	<	2.30	$&$	<	-0.22	$&$	<	+2.95	$&$	0.34	$&	Spec. Synthesis	\\
\ion{Sr}{2}	&$	2	$&$		-0.90	$&$		-3.77	$&$		-0.94	$&$	0.33	$&	Spec. Synthesis	\\
\ion{Y}{2}	&$	4	$&$	<	-0.06	$&$	<	-2.27	$&$	<	+0.94	$&$	0.38	$&	Spec. Synthesis	\\
\ion{Zr}{2}	&$	4	$&$	<	0.80	$&$	<	-1.78	$&$	<	+1.39	$&$	0.34	$&	Spec. Synthesis	\\
\ion{Mo}{2}	&$	1	$&$	<	0.62	$&$	<	-1.26	$&$	<	+1.88	$&$	0.31	$&	Spec. Synthesis	\\
\ion{Ba}{2}	&$	3	$&$		-1.75	$&$		-3.93	$&$		-1.10	$&$	0.33	$&	Spec. Synthesis	\\
\ion{La}{2}	&$	5	$&$	<	-0.18	$&$	<	-1.28	$&$	<	+1.87	$&$	0.32	$&	Spec. Synthesis	\\
\ion{Ce}{2}	&$	5	$&$	<	-0.45	$&$	<	-2.03	$&$	<	+1.10	$&$	0.30	$&	Spec. Synthesis	\\
\ion{Pr}{2}	&$	4	$&$	<	-0.94	$&$	<	-1.66	$&$	<	+1.48	$&$	0.31	$&	Spec. Synthesis	\\
\ion{Nd}{2}	&$	6	$&$	<	-0.50	$&$	<	-1.92	$&$	<	+1.22	$&$	0.31	$&	Spec. Synthesis	\\
\ion{Sm}{2}	&$	5	$&$	<	-0.50	$&$	<	-1.46	$&$	<	+1.69	$&$	0.32	$&	Spec. Synthesis	\\
\ion{Eu}{2}	&$	4	$&$	<	-1.79	$&$	<	-2.31	$&$	<	+0.91	$&$	0.39	$&	Spec. Synthesis	\\
\ion{Gd}{2}	&$	3	$&$	<	-0.03	$&$	<	-1.10	$&$	<	+2.05	$&$	0.32	$&	Spec. Synthesis	\\
\ion{Tb}{2}	&$	3	$&$	<	-0.47	$&$	<	-0.77	$&$	<	+2.41	$&$	0.35	$&	Spec. Synthesis	\\
\ion{Dy}{2}	&$	3	$&$	<	-0.70	$&$	<	-1.80	$&$	<	+1.37	$&$	0.34	$&	Spec. Synthesis	\\
\ion{Er}{2}	&$	2	$&$	<	-0.37	$&$	<	-1.29	$&$	<	+1.97	$&$	0.43	$&	Spec. Synthesis	\\

\label{abundance_table}
\end{deluxetable}

\begin{deluxetable*}{lcccccc}
\tablecaption{Summary of Error Analysis}
\fontsize{8}{6}\selectfont
\tablehead{ \colhead{Species} & \colhead{N} &  \colhead{$\sigma$} & \colhead{$\Delta\textrm{log}_{10}\left(\epsilon_{X}\right)$} & \colhead{$\Delta\textrm{log}_{10}\left(\epsilon_{X}\right)$} & \colhead{$\Delta\textrm{log}_{10}\left(\epsilon_{X}\right)$}&  \colhead{${\Delta \textrm{log}_{10}\left(\epsilon_{X}\right)_{\textrm{,Total}}}$} \\ 
&&&$(\Delta T = \newline +100\textrm{K})$ & $(\Delta \textrm{log}(g) = \newline +0.2\textrm{ dex})$ & $(\Delta \xi = \newline +0.5\textrm{ km s}^{-1})$ &
}
\multicolumn{7}{c}{DES J025540-540807}\\
\tableline
Mg I  	&	2	&	0.13	& $	+0.14	$ & $	-0.08	$ & $	-0.16	$ & $	0.25	$	\\
Si I  	&	4	&	 0.25 	& $	+0.25	$ & $	+0.25	$ & $	+0.25	$ & $	0.45	$	\\
Ca I  	&	4	&	0.03	& $	+0.08	$ & $	-0.03	$ & $	-0.12	$ & $	0.15	$	\\
Sc II  	&	1	&	 0.25 	& $	+0.25	$ & $	+0.25	$ & $	-0.10	$ & $	0.44	$	\\
Ti I  	&	3	&	0.25	& $	-0.25	$ & $	+0.13	$ & $	-0.25	$ & $	0.40	$	 \\
Cr I  	&	8	&	 0.20 	& $	-0.10	$ & $	-0.10	$ & $	-0.25	$ & $	0.30	$	\\
Mn I  	&	3	&	 0.50 	& $	+0.25	$ & $	+0.25	$ & $	+0.50	$ & $	0.61	$	 \\
Fe I  	&	12	&	0.20	& $	+0.11	$ & $	-0.01	$ & $	-0.05	$ & $	0.13	$	 \\
Fe II  	&	4	&	0.13	& $	+0.05	$ & $	+0.07	$ & $	-0.01	$ & $	0.11	$	 \\
Ni I  	&	2	&	 0.25 	& $	+0.25	$ & $	+0.10	$ & $	-0.25	$ & $	0.41	$	 \\
Ba II  	&	3	&	 0.25 	& $	-0.25	$ & $	-0.25	$ & $	-0.25	$ & $	0.46	$	\\
Eu II  	&	2	&	 0.25 	& $	-0.25	$ & $	-0.25	$ & $	-0.10	$ & $	0.41	$	 \\

\tableline
\multicolumn{7}{c}{\rule{0pt}{2ex}DES J025543-544349}\\
\tableline
Mg I  	&	3	&	0.09	& $	+0.09	$ & $	-0.06	$ & $	-0.22	$ & $	0.25	$	\\
Si I  	&	4	&	 0.25 	& $	+0.25	$ & $	+0.25	$ & $	+0.25	$ & $	0.45	$	\\
Ca I  	&	3	&	0.04	& $	+0.09	$ & $	-0.03	$ & $	-0.08	$ & $	0.13	$	\\
Sc II  	&	1	&	 0.25 	& $	+0.25	$ & $	+0.25	$ & $	+0.25	$ & $	0.50	$	 \\
Ti I  	&	3	&	0.25	& $	-0.25	$ & $	-0.13	$ & $	-0.25	$ & $	0.40	$	 \\
Cr I  	&	8	&	 0.20 	& $	-0.10	$ & $	-0.13	$ & $	-0.25	$ & $	0.31	$	\\
Mn I  	&	3	&	 0.50 	& $	+0.25	$ & $	+0.25	$ & $	+0.50	$ & $	0.68	$	 \\
Fe I  	&	12	&	0.29	& $	+0.13	$ & $	-0.02	$ & $	-0.04	$ & $	0.16	$	 \\
Fe II  	&	4	&	0.17	& $	-0.02	$ & $	+0.07	$ & $	-0.01	$ & $	0.11	$	 \\
Ni I  	&	2	&	 0.25 	& $	+0.25	$ & $	-0.25	$ & $	-0.25	$ & $	0.47	$	 \\
Ba II  	&	3	&	 0.25 	& $	-0.10	$ & $	-0.10	$ & $	-0.25	$ & $	0.32	$	 \\
Eu II  	&	2	&	 0.25 	& $	-0.25	$ & $	-0.25	$ & $	-0.25	$ & $	0.47	$	 \\
\tableline
\multicolumn{7}{c}{\rule{0pt}{2ex}DES J025535-540643}\\
\tableline
\ion{C}{0} (CH)	&$	1	$&$	0.25	$&$	+0.15	$&$	-0.05	$&$	-0.20	$&$	0.36	$\\
\ion{N}{0} (CN)	&$	1	$&$	0.25	$&$	+0.25	$&$	+0.25	$&$	+0.25	$&$	0.50	$\\
\ion{Mg}{1}  	&	4	&	0.27	& $	+0.14	$ & $	-0.08	$ & $	-0.16	$ & $	0.30	$	\\
\ion{Al}{1}	&$	2	$&$	0.07	$&$	+0.15	$&$	-0.08	$&$	+0.13	$&$	0.22	$\\
\ion{Si}{1}	&$	1	$&$	0.25	$&$	+0.30	$&$	+0.20	$&$	+0.20	$&$	0.48	$\\
\ion{Ca}{1}  	&	4	&	0.29	& $	+0.08	$ & $	-0.03	$ & $	-0.12	$ & $	0.21	$	 \\
\ion{Sc}{2}	&$	3	$&$	0.12	$&$	+0.10	$&$	+0.08	$&$	-0.05	$&$	0.15	$\\
\ion{Ti}{1}	&$	3	$&$	0.09	$&$	+0.10	$&$	-0.10	$&$	-0.10	$&$	0.18	$\\
\ion{V}{1}	&$	1	$&$	0.25	$&$	+0.10	$&$	+0.10	$&$	+0.10	$&$	0.30	$\\
\ion{Cr}{1}	&$	5	$&$	0.35	$&$	-0.09	$&$	-0.24	$&$	-0.23	$&$	0.38	$\\
\ion{Mn}{1}	&$	1	$&$	0.25	$&$	+0.05	$&$	-0.20	$&$	-0.15	$&$	0.36	$\\
\ion{Fe}{1}  	&	60	&	0.20	& $	+0.11	$ & $	-0.01	$ & $	-0.05	$ & $	0.12	$	 \\
\ion{Fe}{2}  	&	4	&	0.33	& $	+0.05	$ & $	+0.07	$ & $	-0.01	$ & $	0.19	$	 \\
\ion{Co}{1}	&$	3	$&$	0.29	$&$	+0.22	$&$	-0.13	$&$	-0.08	$&$	0.32	$\\
\ion{Ni}{1}	&$	3	$&$	0.47	$&$	+0.17	$&$	+0.12	$&$	+0.08	$&$	0.35	$\\
\ion{Cu}{1}	&$	3	$&$	0.25	$&$	+0.15	$&$	+0.10	$&$	-0.10	$&$	0.32	$\\
\ion{Zn}{1}	&$	2	$&$	0.25	$&$	+0.10	$&$	+0.10	$&$	+0.10	$&$	0.30	$\\
\ion{Ga}{1}	&$	1	$&$	0.25	$&$	+0.20	$&$	+0.20	$&$	+0.20	$&$	0.43	$\\
\ion{Rb}{1}	&$	2	$&$	0.25	$&$	+0.20	$&$	+0.10	$&$	+0.05	$&$	0.34	$\\
\ion{Sr}{2}	&$	2	$&$	0.25	$&$	+0.13	$&$	-0.08	$&$	-0.15	$&$	0.33	$\\
\ion{Y}{2}	&$	4	$&$	0.25	$&$	+0.20	$&$	+0.20	$&$	+0.05	$&$	0.38	$\\
\ion{Zr}{2}	&$	4	$&$	0.25	$&$	-0.20	$&$	+0.05	$&$	+0.10	$&$	0.34	$\\
\ion{Mo}{2}	&$	1	$&$	0.25	$&$	+0.15	$&$	-0.10	$&$	+0.05	$&$	0.31	$\\
\ion{Ba}{2}	&$	3	$&$	0.52	$&$	+0.10	$&$	+0.07	$&$	-0.07	$&$	0.33	$\\
\ion{La}{2}	&$	5	$&$	0.25	$&$	+0.15	$&$	+0.10	$&$	+0.10	$&$	0.32	$\\
\ion{Ce}{2}	&$	5	$&$	0.25	$&$	+0.10	$&$	-0.10	$&$	-0.10	$&$	0.30	$\\
\ion{Pr}{2}	&$	4	$&$	0.25	$&$	+0.15	$&$	+0.10	$&$	+0.05	$&$	0.31	$\\
\ion{Nd}{2}	&$	6	$&$	0.25	$&$	+0.15	$&$	-0.10	$&$	-0.05	$&$	0.31	$\\
\ion{Sm}{2}	&$	5	$&$	0.25	$&$	+0.15	$&$	-0.10	$&$	-0.10	$&$	0.32	$\\
\ion{Eu}{2}	&$	4	$&$	0.25	$&$	-0.20	$&$	-0.20	$&$	-0.10	$&$	0.39	$\\
\ion{Gd}{2}	&$	3	$&$	0.25	$&$	-0.15	$&$	-0.10	$&$	-0.10	$&$	0.32	$\\
\ion{Tb}{2}	&$	3	$&$	0.25	$&$	-0.10	$&$	-0.20	$&$	-0.10	$&$	0.35	$\\
\ion{Dy}{2}	&$	3	$&$	0.25	$&$	-0.15	$&$	-0.15	$&$	-0.10	$&$	0.34	$\\
\ion{Er}{2}	&$	2	$&$	0.25	$&$	-0.20	$&$	-0.20	$&$	-0.20	$&$	0.43	$\\

\label{abundance_error_table}
\end{deluxetable*}

\begin{deluxetable}{lcc}
\tablecaption{Supernova Yield Model Fits to DES J025535-540643} 
\tablehead{\colhead{Model} & \colhead{Best Fit} & \colhead{Mean } \\ 
&Progenitor Mass&Sq. Residual}
O Shell ($S=4$) Piston& $10\ M_\sun$ & 23.8  \\
Fe Core ($Y_e$) Piston& $85\ M_\sun$ & 28.2  \\
PISN & $260\ M_\sun$ & 64.4\\
\label{sn_yield_model_table}
\end{deluxetable}

\end{document}